\documentclass[fleqn,usenatbib,useAMS]{mnras}
\usepackage{amsmath} 
\usepackage{graphicx}
\usepackage{amssymb}

\def \prd {PRD}
\def \apj {ApJ}
\def \apjs {ApJS}
\def \apjl {ApJL}
\def \mnras {MNRAS}
\def \aap {A\&A}

\def \araa {ARAA}

\def \nat {Nature}

\def\t9{T$_{90}$}

\newcommand{\urate}{$\mathrm{Gpc}^{-3}\, \mathrm{yr}^{-1}$}

\title[Host galaxies of binary black hole mergers]{Host galaxy identification for binary black hole mergers with long baseline gravitational wave detectors.}

\author[Howell, Chan, Chu et al.]
{E. J. Howell$^1$\thanks{E-mail:eric.howell@uwa.edu.au}, M. L. Chan$^2$, Q. Chu$^1$, D. H. Jones, I. S. Heng$^2$, H. -M. Lee$^4$, \newauthor D. Blair$^1$,J. Degallaix$^5$, T. Regimbau$^6$, H. Miao$^7$, C. Zhao$^1$, M. Hendry$^2$, D. Coward$^1$,
   \newauthor  C. Messenger$^2$, L. Ju$^1$, Z.-H. Zhu $^{8,9}$ \\
\\
$^1$ OzGrav-UWA, School of Physics and Astrophysics, University of Western Australia, Crawley WA 6009, Australia\\
$^2$ SUPA, School of Physics and Astronomy, University of Glasgow, Glasgow G12 8QQ, United Kingdom\\
$^3$ English Language and Foundation Studies Centre, University of Newcastle, Callaghan NSW 2308, Australia\\
$^4$ Seoul National University, Seoul 08826, Korea\\
$^5$ Laboratoire des Mat$\acute{e}$riaux Avanc$\acute{e}$s (LMA), CNRS/IN2P3, F-69622 Villeurbanne, France\\
$^6$ Artemis, Universit$\acute{e}$ C$\hat{o}$te d'Azur, Observatoire C$\hat{o}$te d'Azur, CNRS, CS 34229, F-06304 Nice Cedex 4, France\\
$^7$ School of Physics and Astronomy, Institute of Gravitational Wave Astronomy, University of Birmingham, United Kingdom\\
$^8$Department of Astronomy, Beijing Normal University, Beijing 100875, China\\
$^9$School of Physics and Technology, Wuhan University, Wuhan 430072, China\\
}

\date{Accepted XXX. Received YYY; in original form ZZZ}

\pubyear{2017}

\begin{document}
\label{firstpage}
\pagerange{\pageref{firstpage}--\pageref{lastpage}}
\maketitle

\begin{abstract}
The detection of three black hole binary coalescence events by Advanced LIGO allows the science benefits of future detectors to be evaluated. In this paper we report the science benefits of one or two 8km arm length detectors based on the doubling of key parameters in an advanced LIGO type detector, combined with realisable enhancements. It is shown that the total detection rate for sources similar to those already detected, would increase to $\sim$ 10$^{3}$--10$^{5}$ per year. Within 0.4Gpc we find that around 10 of these events would be localizable to within $\sim 10^{-1}$ deg$^2$. This is sufficient to make unique associations or to rule out a direct association with the brightest galaxies in optical surveys (at r-band magnitudes of 17 or above) or for deeper limits (down to r-band magnitudes of 20) yield statistically significant associations. The combination of angular resolution and event rate would benefit precision testing of formation models, cosmic evolution and cosmological studies.
\end{abstract}



\section{Introduction}

The first gravitational wave (GW) observations of the coalescence and merger of stellar mass binary black holes (BBHs) by Advanced LIGO \citep[aLIGO;][]{aLIGO_2015CQGra} prove the existence of a large cosmic population  of these sources \citep{O1_BBHs_2016,2016PhRvL_GW150914,2016PhRvL_GW151226}. BBHs could be the relics of co-evolved binaries from the Pop III or Pop II era \citep[as proposed by][]{Lipunov1997, Dominik2015ApJ,Belczynski_2016_Natur}, or could be dynamically formed in globular clusters \citep{Bae2014MNRAS,Hong_2015MNRAS,Rodriguez2015,Rodriguez2016,Rodriguez2016ApJ}.

The GW observations of each event provide information on the masses and spins of the BBH system, and their location in a spatial volume element determined by the angular resolution and the luminosity distance error. However, the \emph{mass-redshift degeneracy}, which results from the intrinsic mass $M$ being redshifted in the observer frame to $M_{\mathrm{z}}=(1 + z)M$, leads to errors in the luminosity distance which for a given cosmological model can be resolved through independent measurement of the redshift of the host galaxy.

To use the BBH population as a tool for distinguishing the formation mechanism, one requires a large population of such events \citep{Stevenson_2015ApJ}. Furthermore, to use GWs to investigate cosmology it would be necessary to be able to identify the host galaxies \citep{Fan2014ApJ}, and measure their redshift. Independent redshift measurements would also significantly improve the accuracy of measuring the mass parameters by breaking the mass-redshift degeneracy \citep{Messenger_2014,Ghosh_PRD_2015}.

Now that GW detectors have crossed the threshold from non-detection to detection, assuming only a roughly homogeneous distribution of sources in the cosmos, we can estimate the number of detectable events as a function of detector sensitivity. This depends on the cosmic event rate density, integrated over the accessible volume of the universe. At low redshift ($z \approx 0.2$ the event rate scales roughly as the cube of detector sensitivity so that the estimates of improved sensitivity can be made using limited assumptions. For higher redshifts, cosmic expansion and the distribution of coalescence times influences the event number count.

The angular resolution of gravitational wave detector networks largely follows the Rayleigh criterion, $ \Delta \Theta = (\lambda/D) \rho^{-1} $, where $D$ is the detector spacing and $\rho$ is the signal-to-noise ratio (SNR), as discussed in detail below. Except for the lowest frequencies, which to date have been difficult to access due to technical noise, detector spacings are a few wavelengths (100\,Hz $\approx$ 3000\,km wavelength). There is a substantial advantage in measuring both the relative phase and the amplitude of incoming signals to determine their position in the sky.  Maximum possible baselines are always beneficial.

The current world array of detectors consist of aLIGO, Advanced Virgo \citep[AdV; expected to be on line in 2017][]{TheVirgo:2014hva}, KAGRA in Japan \citep{KAGRA_PhysRevD_2013}, expected to contribute data a few years later, and LIGO-India\footnote{\mbox{see \url{https://dcc.ligo.org/LIGO-M1100296/public} for Detailed} Project Report} has been proposed for 2022. The detector network noticeably omits a southern hemisphere detector, even though such a detector would contribute long baselines to all the northern detectors \citep{WenChen2010PhRvD,Klimenko2011PhRvD,Raffai2013CQGra, Chu2016MNRAS}.

There have been several proposals for third generation gravitational wave detectors (3G) \citep{Punturo_ET_2010,Dwyer2015PhRvD,Abbott_CE_2016}. Options include extending the arms from 4km to lengths of the order 10km to 40km \citep{Dwyer2015PhRvD,Abbott_CE_2016}, the use of cryogenics with silicon test masses \citep[][]{Punturo_ET_2010} and optics operating at about 2 microns wavelength \citep{Abbott_CE_2016}. Such detectors are likely to come on line in the late 2020s, but necessarily depend on substantial research and development.

To bridge the gap between present second generation (2G)and future 3G interferometric GW observatories (IFOs) an 8km LIGO-type 2.5G detector has been proposed \citep{Blair2015SCPMA} that would be constructed by scaling up currently proven technology. The 8km arm length was chosen as a safe length for which LIGO-type stainless steel vacuum pipes could be used; pipe diameter is a major cost driver. As LIGO was initially designed to house up to three IFOs, doubling the arm length while maintaining the same pipe diameter allows implementation using the LIGO design.

The design also included doubling the fused silica suspension fibre length, (to 1.5m) and a doubling of the test masses to 80kg. Two other enhancements match proposals for the enhancement of Advanced LIGO: the use of Newtonian noise suppression and squeezed light technology. These technologies are already under development and could be tested and implemented in parallel with detector construction.

In this paper we analyse the angular resolution and source event rate that could be expected if one or two 2.5G detectors were added to the existing and planned worldwide array. We show that a detector in China increases the event rate and also improves the angular resolution. However an optimal arrangement is for this detector to be complemented by a southern hemisphere detector; we find that if only one detector were to be installed, an Australia based instrument is the optimal choice.

We will show that of order $\approx 10^{4}$ detections per year could be expected in the improved network – corresponding to one event every hour. The strongest sources at about 0.4\,Gpc have a SNR $\sim 100$, and can be localized sufficiently (both in direction and redshift) that the brightest host galaxies of $\approx$1-10 sources per year could be directly identified without confusion. At larger ranges the uncertainty value (measured in sky area $\times$ redshift interval) expands steeply due to: a) the distance dependence of the linear size of the error ellipse; b) the increased angular error and luminosity distance error as the SNR decreases. This causes rapidly increasing source confusion and thus the number of candidate galaxies increases.

The detection prospects of other important GW sources, such as stochastic backgrounds, binary neutron star coalescence, and continuous waves from rotating neutron stars could be enhanced by 2.5G instruments; in this study we choose to base our analysis on only the known BBH sources which have crossed the detection threshold.

The structure of this paper is as follows: Section 2 presents a conceptual outline of an 8km GW interferometric detector and Section 3 presents the angular resolution improvement obtained by adding such a detector to existing and proposed networks. The expected event rates are calculated in Section 4 and the possibility for host galaxy identification is assessed in Section 5. Finally, Section 6 reviews the results and discusses the possibility of building 8km detectors.

\section{Design concept for a 8km arm-length GW interferometric detector}

\begin{figure}
 \centering
 \includegraphics[scale = 0.5,origin=rl]{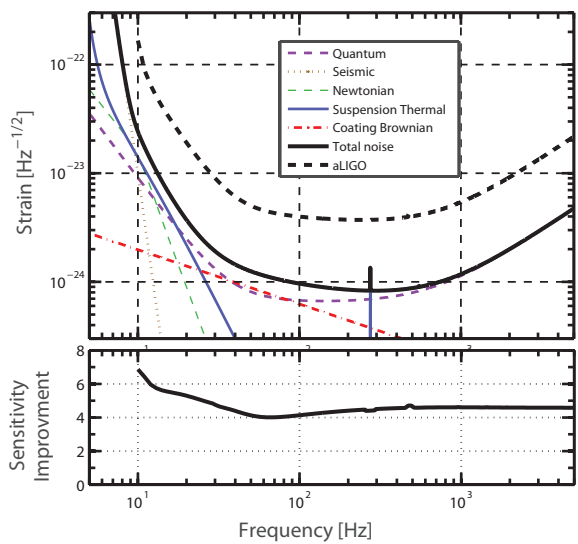}
 \caption{The noise budget of the 8km GW detector described in this paper is shown in the top panel. The instrument has an 8km arm length, 80kg test masses and 250W input laser power; the beam spot radius on input and end test masses are 11cm and 12cm respectively. The detector incorporates a 1.5m test mass last stage suspension and Newtonian noise cancellation. Frequency dependent squeezed vacuum is generated by passing the 8dB phase squeezed vacuum through a 100m filter cavity. For reference an aLIGO design sensitivity noise curve is shown. The lower panel shows the sensitivity improvement of the 8km design over aLIGO.}
  \label{fig:detector}
\end{figure}

Gravitational wave detectors must achieve an optimum balance of noise sources which include quantum noise, thermal noise, residual gas noise, Newtonian noise and seismic noise. The performance of LIGO and aLIGO has demonstrated a good current understanding of detector noise, although at low frequency some noise sources are not fully understood. Since GWs induced relative displacements increase proportional to arm length, sensitivity should also increase proportional to arm length. Effectively, local noise sources associated with the test masses are diluted by increasing the arm length.

Further improvements can also be achieved by increasing the dimensions of the test mass system. Firstly, by increasing the test mass size and mass, one can achieve improved averaging of the thermal noise due to internal acoustic modes. Secondly, by lengthening the test mass suspension fibres, the pendulum frequency of the suspension is reduced, which acts to reduce the pendulum mode thermal noise.

The dominant noise budget and the total detector noise that results from adopting these and other improvements is shown in Figure \ref{fig:detector} \citep[see also][]{Blair2015SCPMA}. The lower panel shows a nearly 4-fold improvement factor. However, this is not a direct result of the doubling of parameters, but a composite effect due to the addition of thermal noise and quantum noise.

A key aspect of this proposed detector would be provision for upgrading from a 2.5G design to 3G technology \citep[such as cryogenic detectors with silicon test masses;][]{Punturo_ET_2010} in the future when these technologies are proven.

The specific parameters used to improve over aLIGO technology and define the above results are: a) an extending of the arm length to 8\,km;\, b) doubling the test mass weight to 80kg;\, c) extending the last stage test mass suspension to 1.5m;\, d) increasing the injected laser power to 250W;\, e) doubling the spot-area;\, f) implementing an 8dB phase squeezed vacuum with a 100m filter cavity to create frequency dependent squeezed vacuum;\, g) 2$\times$ Newtonian noise cancellation. We briefly address the feasibility of these proposed improvements below.

The test masses forming the long arm cavities are the most critical optics of the interferometer. The arm cavity is the transducer which impinges the gravitational waves signal to the phase of the light which can then be measured through interference. Any imperfection on the test masses can absorbed or diffused by the light, thus creating excessive optical loss and ultimately reducing the sensitivity of the detector.

Due to the laser beam diffraction along its propagation over 8\,km, the test masses must have a large diameter to be able to reflect all the laser light. The cylindrical optics will have a diameter of 500~mm and weigh 80\,kg. Since the mirrors can support a large laser beam, it presents several advantages: the thermal noise can be lowered due to improved averaging over the mirror surface and the decreased laser power density mitigates thermal effects induced by the optical absorption.

Test mass substrates would be made with the purest fused silica: Heraeus Suprasil 3002. This material is available in large size and presents outstanding optical and mechanical properties at room temperature; hence it is already used for current second generation
interferometers \citep{Mitrofanov_2015,Pinard:16}.

The polishing to shape the mirror profile would be specified at the nanometer level, using ion beam figuring. This technology guarantees excellent surface accuracy and minimal roughness, thus limiting the amount of
scattered light \citep{Pinard_2016_IBS}.

The test masses will have a high reflective side (inside the arm cavity) whereas the second side will receive an antireflection coating. Both treatments will be made with the Ion Beam Sputering (IBS) technology to ensure very low optical loss and also low thermal noise level \citep{Pinard_2016_IBS}. It has been confirmed that the LMA 'Grand Coater' machine in France is able to coat test masses to our planned dimensions.

As discussed earlier, the proposed implementation here is all based on proven technologies and hence carries no substantial risk. As an example, a 550~mm diameter beamsplitter has already been produced and installed in the AdV interferometer.

To achieve a full factor of four strain sensitivity improvement over the full aLIGO frequency band requires two additional technologies. The first is quantum squeezing. This technology has already been demonstrated on gravitational wave detectors at high frequency band \citep{Goda2008,AasiJ.2013}, but frequency-dependent squeezing is required to improve the quantum noise limited sensitivity over the full detection band to achieve the above design goal \citep{PhysRevD.65.022002,PhysRevLett.116.041102}. Two critical issues are the optical losses in the interferometer input and output optical path, and the optical loss of the filter cavity. Using the parameters of aLIGO upgrade (A+) \footnote{\url{https://dcc.ligo.org/public/0113/T1400316/004/T1400316-v5.pdf}}, Fig.\ref{fig:detector} shows that the the target sensitivity is achievable.

Finally at the lowest frequencies, it is expected that Newtonian noise associated with varying local gravitational forces acting on the test masses could create additional noise. This noise can be suppressed by measuring the gravitational sources such as passing low frequency atmospheric or ground (seismic) waves and compensating by subtraction \citep{0264-9381-33-23-234001,0264-9381-33-24-244001,PhysRevD.92.022001}. The design proposed for a 8\,km detector assumes Newtonian noise suppression by 2$\times$, which is a modest suppression factor comparing to that proposed for other detector designs, such as LIGO Voyager$^{2}$.

\section{Event rates }
The first detections of GWs from BBH mergers at $z \approx 0.1$ firmly placed these events in the cosmological regime. For aLIGO, the horizon volume in which a GW150914 type event could have been detected during O1 extends out to $z \simeq 0.4$ \citep{GW150914_AstImp_2016ApJ} and at design sensitivity extends to around $z = 1$; Third Generation detectors such as ET should routinely detect such events out to $z \sim 10$ and beyond \citep{Vitale2016PhRvD,Vitale2017PhRvD}. The ranges accessible by an 8km IFO, and indeed upgraded configurations of aLIGO, will extend far beyond the Euclidean regime. For these future instruments, projected estimations of the number counts of detected sources require a more rigorous treatment of cosmic source rate evolution. Additionally, in comparison with lower mass systems which can be reasonably modelled by consideration of only the inspiral part of the signal, higher mass systems that merge at lower frequency and have less cycles in the detector band require detailed modelling of the inspiral and ringdown phases.

In the following sections we will present a working framework to model the cosmic source rate evolution and to calculate the astrophysical reach of gravitational wave IFOs to BBH mergers. At the present early stage of GW astronomy there are many uncertainties in the details of the BBH populations e.g the mass distributions, the intrinsic event rates and their evolutionary pathways. We therefore produce estimates based on confirmed events assuming two canonical sources; 30--30 $M_{\odot}$ representing heavy events (such as GW150914 and GW170104 with total masses 65$M_{\odot}$ and 50$M_{\odot}$ respectively) and 10--10 $M_{\odot}$ for lower mass systems (e.g. GW151226 with a total system mass of 22$M_{\odot}$). We assume the same source rate evolution and event rate ranges for both these canonical populations.

\subsection{The all sky event rate equation of BBH coalescence}
\label{sect:event_rate}
To estimate the number of coalescing BBHs accessible by an 8km IFO we firstly assume that their formation rate $R_{\mathrm{BBH}}(z)$ tracks the star formation history of the Universe with a delay time $t_{\mathrm{d}}$ from formation until final merger. As heavy BBH systems such as GW150914 and GW170104 are predicted to form
in low-metallicity stellar environments we also include a measure cosmic metallicity dependence.

For the star formation history of the Universe $R_{\star}(z)$ (in $\mathrm{M}_{\odot}\,\mathrm{yr}^{-1}\, \mathrm{Mpc}^{-3}$) we adopt the extinction-corrected cosmic star formation rate model of \citet{MadauDickinsonSFR2014ARA&A}, which combines measurements from sources observed at ultraviolet and far-infrared.

To estimate metallicity evolution we adopt the model of \citet{langerNorman_metalisity_06} which estimates the fraction $\Psi(z,\xi)$ in redshift $z$ of massive stars with metallicities less than some cutoff value $\xi = Z/Z_{\odot}$. Following \citet{GW150914_AstImp_2016ApJ,BBH_GW150914_SGWB_2016PhRvL} we adopt a value of $\xi = 0.5$.
\begin{figure}
 \centering
 \includegraphics[bbllx = 70pt,bblly =5pt, bburx = 350pt, bbury =280pt,scale = 0.65,origin=rl]{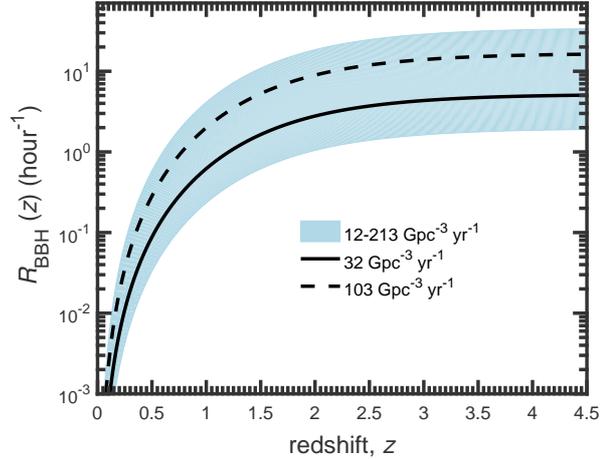}
 \caption{The cumulative all sky rate of BBH coalescence based on the the rate estimates of \citet{GW:170104}for the 4 events: GW150914, LVT151012, GW151226 and GW170104. The solid and dashed curves show the median values determined using a power-law and in-log uniform distributions of masses respectively; the shaded region shows the combined rate range. The plot shows that the Universal rate of BBH mergers is of order 10s/hour.
 }
  \label{fig:dr}
\end{figure}
A model of cosmic BBH formation rate including a metallicity dependence is obtained by scaling the star formation history, $R_{\star}(z)$, with the function $\Psi(z, \xi)$:
\begin{equation}\label{eq_formation_rate}
    R_{\rm F}(z)= \Psi(z, \xi) R_{\star}(z)\,.
\end{equation}
To model the cosmic BBH merger rate $R_{\mathrm{BBH}}(z)$, one must account for the delay time $t_{\rm d}$  between formation $t_{\rm f}$ and the age of the Universe at the time of merger $t(z)$ \citep{Zhu_BBH_SGWB_2011ApJ}. If $z_{\rm f}$ and $z$ represent the redshifts at which BBH systems form and merge respectively, the delay time for BBHs is given by:
\begin{equation}
t_{d}= \int_{z}^{z_{\mathrm{f}}} \frac{\mathrm{d}z} {(1+z')H(z')}.
\end{equation}
\noindent The BBH merger rate is calculated by convolving the BBH formation rate $R_{\rm F}(z)$ with the delay time distribution $P(t_{\rm d})$ \citep{Regimbau2009PhRvD}:
\begin{equation}\label{eq_bbh_rate}
    R_{\mathrm{BBH}}(z)= \int_{t_{\mathrm{min}}}^{t_{\mathrm{max}}} R_{\mathrm{F}}(z_{\mathrm{f}}) P(t_{\mathrm{d}}) \mathrm{d}t
\end{equation}
\noindent Following \citet{BBH_GW150914_SGWB_2016PhRvL} we take, $P(t_{\rm d}) =1/t_{d}$, for $t_{\rm d} > t_{\rm min}$, where $t_{\rm min} = 50$\,Myr is the minimum delay time for a BBH system to evolve to merger and $t_{\rm min}$ is equal to the Hubble time.

The differential BBH merger rate, $\mathrm{d}R_{\mathrm{BBH}} /\mathrm{d}z$, which describes the event rate within the redshift shell $z$ to $z+{\mathrm d}z$ is then:
\begin{equation}\label{eq_drdz}
\mathrm{d}R_{\mathrm{BBH}} = \frac{\mathrm{d}V}{\mathrm{d}z} \frac{e(z)\, r_{0}}{1+z} \mathrm{d}z \,,
\end{equation}

\noindent with $e(z)$ a source rate evolution model formed by normalising the BBH merger rate, $R_{\rm BBH}(z)$, to unity in the local Universe $(z = 0)$; this allows a scaling by the event rate of BBH mergers $r_{\mathrm{0}}$ (volumetric per unit time). The $(1 + z)$ factor accounts for time dilation of the observed rate through cosmic expansion, converting a source-count to an event rate. The co-moving volume element:
\begin{equation}\label{dvdz}
\frac{\mathrm{d}V}{\mathrm{d}z}= \frac{4\pi c}{H_{0}}\frac{d_\mathrm{\hspace{0.25mm}L}^{\hspace{1.5mm}2}(z)}{(1 +
z)^{\hspace{0.25mm}2}\hspace{0.5mm}h(z)}\,,
\end{equation}
\noindent describes how the number densities of non-evolving objects locked into Hubble flow are constant with redshift. The quantity $d_{\mathrm{L}}(z)$ is the luminosity distance \citep[cf.][p.\hspace{0.5mm}332]{Peebles}. For a `flat-$\Lambda$' cosmology, we employ the cosmological parameters $\Omega_{\mathrm
m}=0.31$, $\Omega_{\mathrm \Lambda}=0.69$ and
\mbox{$H_{0}=67.8$ km s$^{-1}$ Mpc$^{-1}$} \citep{Planck2015}.

A cumulative event rate of BBH mergers throughout the Universe is estimated by integrating equation (\ref{eq_drdz}):
\begin{equation}
R_{\mathrm{BBH}}(z) = \int_{0}^{z}(\mathrm{d}R_{\mathrm{BBH}} /\mathrm{d}z)\mathrm{d}z\;,\label{eq_Rz}
\end{equation}
\noindent Figure \ref{fig:dr} shows the function $R_{\mathrm{BBH}}(z)$ based on the rates given in \citet{GW:170104} \citep[updated from ][]{O1_BBHs_2016} for the 4 events: GW150914, LVT151012, GW151226 and GW170104. The solid and dashed curves show the median values determined using a power-law ($103^{+110}_{-63}$ \urate)and in-log uniform distribution($32^{+33}_{-20}$ \urate) of masses respectively; the shaded region shows the combined rate range for the two distributions of 12--213 \urate.

The plot shows that the rate is asymptotic from about $z=3$; this is due to a peak in the differential rate $\mathrm{d}R_{\mathrm{BBH}} /\mathrm{d}z$ occurring at around $z \approx 1.5$. The Universal rate of BBH mergers based on this estimation is of order 10s min$^{-1}$ or approximately one event every 5 minutes. To estimate source counts one must fold the function $R_{\mathrm{BBH}}(z)$ with a redshift dependent detection efficiency function. We will first consider the optimal case which is a horizon range for optimally located and orientated sources.

\subsection{The detection ranges for BBH coalescenses}
\label{sect:astro_reach}
The horizon redshift $z_{\mathrm{H}}$, is the maximum redshift for which an optimally orientated and located source can be detected for some sensitivity threshold or SNR. To calculate this quantity we use equation B11 of \citet{Ajith_2008} which provides the optimal SNR, $\rho_{\mathrm{opt}}$ as a function of distance for a non-spinning phenomenological inspiral-merger-ringdown BBH waveform in the frequency domain, for a given GW detector sensitivity noise curve. This equation is parameterised in terms of the total mass $M=m_{1}+m_{2}$ and reduced mass $\eta =m_{1} m_{2}/M^{2}$ using a set of coefficients \citep[See Table 1 in][]{Ajith_2008}. The SNR is determined through the redshifted quantities $M_{z} = M(1 + z)$ and $\eta_{z} = \eta(1 + z)$; values of $z_{\mathrm{H}}$ are calculated iteratively through $z$ by determining the components $M_{z}$ and $\eta_{z}$ required to produce a given $\rho_{\mathrm{opt}}$.

\begin{figure}
 \centering
 \includegraphics[bbllx = 70pt,bblly =-10pt, bburx = 350pt, bbury =280pt,scale = 0.55,origin=rl]{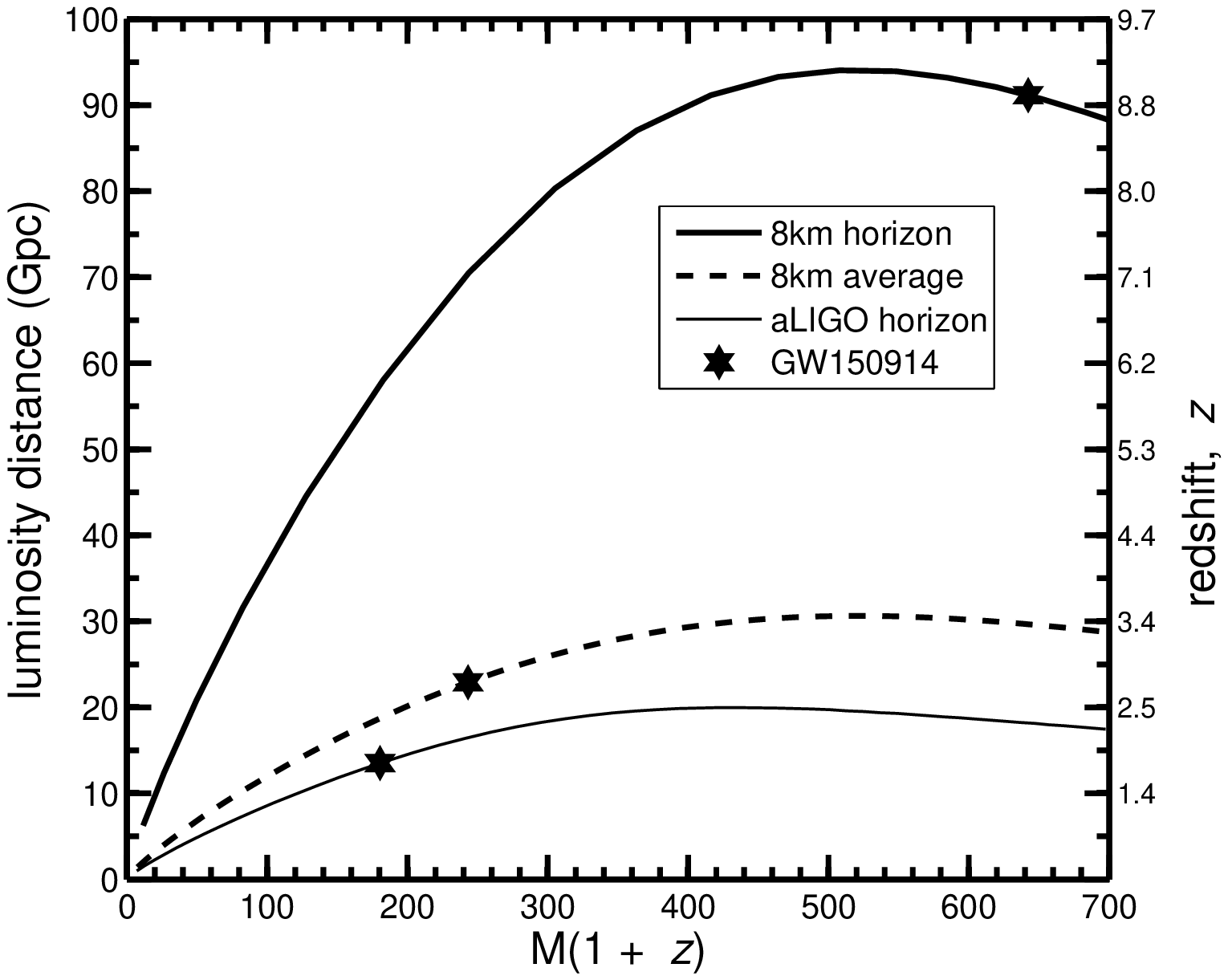}
 \caption{The astrophysical reach for BBH mergers as a function of the observed (redshifted) total system mass assuming non-spinning and equal mass components. The solid lines show the maximum observable distances (in Gpc and cosmological redshift) for an 8-km (thick line) and an aLIGO IFO \citep[thin line; this is consistant with Fig 2 of][]{Ghosh_PRD_2015}. The thick dashed line shows an average distance assuming a 50\% probability of detection. The stars indicate the positions on each curve for a GW150914 type event).
 }
  \label{fig:horizon}
\end{figure}

Figure \ref{fig:horizon} illustrates the astronomical reach of an 8-km IFO for different BBH systems of equal mass components and total redshifted masses $M_{z}$. The ranges $z_{\mathrm{H}}$ are calculated using a value of $\rho_{\mathrm{opt}}$=8 which following standard convention can act as a proxy for a detector network \citep{Abadie2010CQG,Stevenson_2015ApJ,GW150914_AstImp_2016ApJ}\footnote{To be 99\% confident that a GW has been detected, an SNR $\sim 8 $ is equivalent to a false-alarm-rate of 1 in 100 years of observation \mbox{($3 \times 10^{-10} $\,Hz )}.}. For comparison, we also calculate the aLIGO ranges using the zero-detuning design sensitivity noise curve\footnote{\url{https://dcc.ligo.org/LIGO-T1200307}}.

The plot illustrates the impressive astronomical reach obtained by an 8km IFO for an optimally oriented and located GW150914 type system (in terms of the estimated intrinsic component masses); the plot shows that the horizon distance for such a system is $z_{\mathrm{H}}=8.9$; the comparative estimate for aLIGO at design sensitivity is $z_{\mathrm{H}}=1.8$. We note that although the definition of $z_{\mathrm{H}}$ is a useful measure of a detectors capabilities, observing optimally orientated and located events will be rare; to provide more realistic detection rate estimates, in the next section we will model the detection efficiency as a function of redshift.

\begin{figure}
 \centering
 \includegraphics[scale = 0.55,origin=rl]{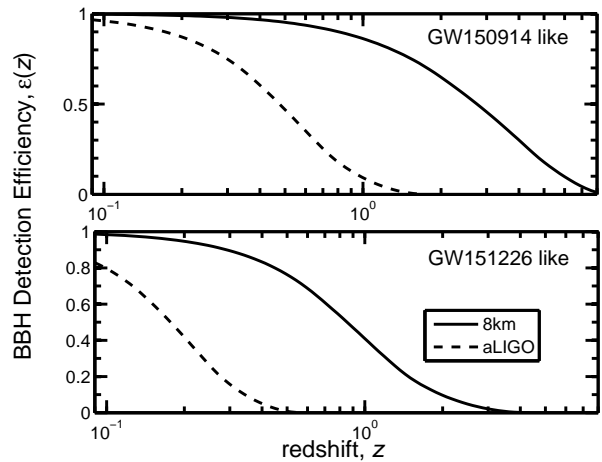}
 \caption{The detection efficiency functions for BBH mergers with total system masses (assuming non-spinning and equal mass components) equal to 30-30 $M_{\odot}$ (top) and 10-10 $M_{\odot}$ BBH mergers (bottom). Curves are shown for both 8km IFOs and aLIGO at design sensitivity. The plot emphasises how lighter systems such as GW151226 ($\sim$ 10-10 $M_{\odot}$) will only be detected within smaller cosmological volumes than heavier GW150914 ($\sim$ 30-30 $M_{\odot}$) type systems.
 }
  \label{fig:efficiency}
\end{figure}

\subsection{The detector efficiency function for BBH inspirals}
\label{section:eff_function}

To estimate of the fraction of sources that exceed a detection threshold as a function of redshift, $\epsilon(z)$, we follow the approach of \citet{Belczynski2014ApJ} and \citet{Dominik2015ApJ} who utilised the \emph{projection parameter} $\omega$ \citep{FinnChernoff1993PhRD}. This quantity describes the detector responses for different values of sky location, inclination and polarisation of a GW source; for an optimal face-on source directly above a GW detector $\omega=1$ and conversely $\omega=0$ for a poorly located and orientated event. A cumulative distribution function of this quantity $c(\omega)$, which contains all the information of single detector response have been provided analytically by both \citet{Finn1996PhRvD} and \citet{Dominik2015ApJ}.

For a system described by the rest frame quantities \{$M$,$\eta$\} and assuming a SNR threshold $\rho_{\mathrm{th}}$=8, for each value of $z$ the corresponding $\rho_{\mathrm{opt}}$ can be calculated through the procedure outlined in section \ref{sect:astro_reach} for each corresponding \{$z$, $M_{z}$\}. The function $\epsilon(z)$ can then be constructed by mapping $\omega=\rho_{\mathrm{th}}/\rho_{\mathrm{opt}}$ to the distribution $c(\omega)$ through:
\begin{equation}\label{eq_cw}
  \epsilon(z) = c(\omega)=c(\rho_{\mathrm{th}}/\rho_{\mathrm{opt}}(M_{z},\eta_{z}))\,,
\end{equation}
\noindent producing a set of efficiency curves as a function of redshift for BBH mergers of different component masses.

Figure \ref{fig:efficiency} shows the efficiency functions for GW signals from both heavy 30--30$M_{\odot}$ and less massive 10--10$M_{\odot}$ type systems composed of non spinning equal mass components. It is quite evident that lighter systems such as GW151226 will only be detected within smaller cosmological volumes than heavier GW150914 type systems. If one assumes both types of source are drawn from the same population, it is reasonable to expect that larger systems, accessible at greater cosmological volumes, will dominate the statistics.

As detections at the horizon will be rare, it is interesting to estimate an average sensitive range for BBH systems \footnote{In a Euclidean regime, the sensitive and horizon volumes obey the scaling $V_{\mathrm{sensitive}}/V_{\mathrm{horizon}} \approx (1/2.26)^{3}$; at cosmological volumes this approximation is no longer valid.}; in this study we define such a quantity as the value of $z$ at which the detection efficiency is equal to 50\%; for a 8km IFO (aLIGO) the sensitive ranges are $z=$2.8(0.49) for 30--30$M_{\odot}$ and $z=$0.86 (0.18) for 10--10$M_{\odot}$ type events.

As a crude verification of our efficiency calibrations one can use the recorded values of GW150914 with the sensitivity curves of the first aLIGO observing run (O1\footnote{\mbox{Using the `early' curves of } \url{https://dcc.ligo.org/LIGO-T1200307}}). Assuming GW150914 was of average orientation \citep[there is posterior support for GW150914 being face off; see][]{O1_BBHs_2016} and location we construct equation \ref{eq_cw} assuming the a threshold SNR of $\rho_{\mathrm{th}} =23.7$ equal to that recorded for GW150914 and determine the redshift at which $\epsilon(z) \approx 50\%$. We find a value of $z \approxeq 0.07$ which is within the published range, $z\sim 0.09^{+0.03}_{-0.04}$ for this event \citep{2016PhRvL_GW150914}.

\subsection{BBH coalescence source counts}
To calculate the source counts for our two canonical populations of BBH mergers; 30--30 $M_{\odot}$ representing heavy events (such as GW150914 and GW170104) and 10--10 $M_{\odot}$ for lower mass systems (e.g. GW151226) we use the frameworks provided in sections \ref{sect:event_rate}-\ref{section:eff_function}. To calculate cumulative event rates we use the range estimates provided in \citet{GW:170104} of 12--213 \urate for upper and lower limits on the total source distribution.

We note that a more accurate estimation of source counts could be produced by, for example, convolving the function $R_{\mathrm{BBH}}(z)$ with the fraction of sources accessible at each $z$, estimated by integrating over the detector accessible fraction of the chirp mass distribution; however, until we accumulate a larger number of detections such a scheme would require many assumptions. We therefore use a simpler scheme based on present knowledge, using our two canonical systems; we suggest that the results provided by the approximations can be provide reasonable upper and lower limits based on: high mass/upper rate [30--30 $M_{\odot}$/ 213 \urate] and lower mass/lower rate [10--10 $M_{\odot}$/ 12 \urate].

\begin{figure}
 \centering
  \includegraphics[scale = 0.65,origin=rl]{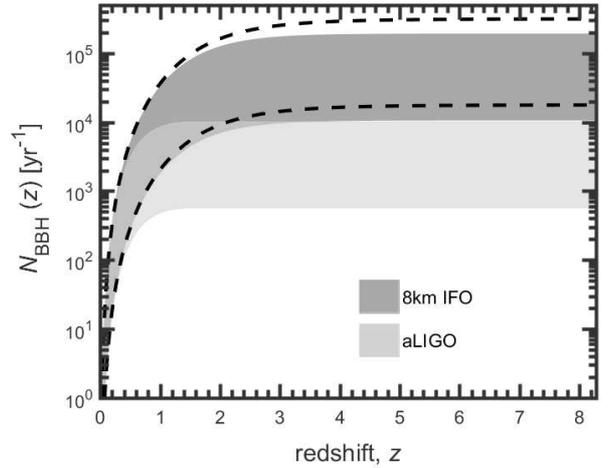}
 \caption{The cumulative detection rate of BBH mergers as a function of redshift based on our modelling for an 8km IFO (bold curve) and an aLIGO IFO at design sensitivity (thin curve). The shaded area indicates the upper and lower limits on the detection rate of 30-30 $M_{\odot}$ mergers based on 4 aLIGO observations. The detection number becomes asymptotic at around $z = 3$ for an 8km IFO and $z = 1$ for aLIGO. The range on the intrinsic number of events (12-213 \urate) are bracketed by the dashed curves.}
  \label{fig:num_GW150914}
\end{figure}

\begin{figure}
 \centering
 \includegraphics[bbllx = 80pt,bblly =5pt, bburx = 350pt, bbury =280pt,scale = 0.65,origin=rl]{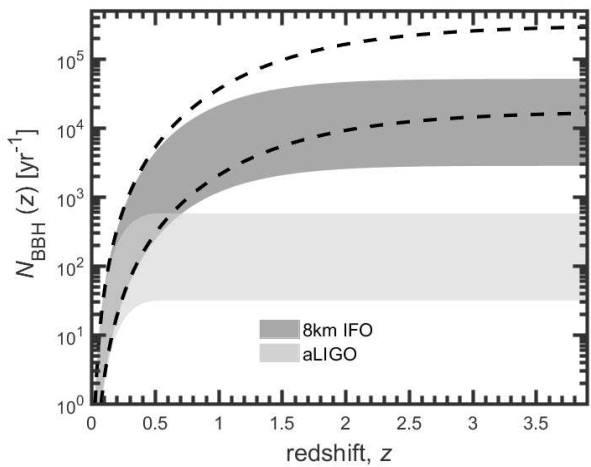}
 \caption{As for Fig.\ref{fig:num_GW150914} but assuming less massive equal mass BBH systems of 10-10 $M_{\odot}$ similar to GW151226. In comparison with 30-30 $M_{\odot}$ type mergers, a smaller fraction of the population are detected due a decreased detection efficiency (see Fig. \ref{fig:efficiency}. }
   \label{fig:num_GW151226}
\end{figure}

Figure \ref{fig:num_GW150914} shows the cumulative number of detections with redshift for both an 8km IFO and aLIGO assuming a population dominated by sources of 30--30 $M_{\odot}$. The detection number becomes asymptotic at around $z = 3$ for an 8km IFO and $z = 1$ for aLIGO; as the differential source rate peaks at around $z \approx 1.5$ there is minimal contribution beyond this range despite the fact that the average reach (50\% detection efficiency) is around $z \approx 2.8$ for an 8km instrument.

The curves show that if all BBH sources were made up of 30--30 $M_{\odot}$ systems, an 8km IFO could detect nearly 10$^{4}$ to 200,000 events each year. This is around 60\% of the intrinsic population shown by the dashed curves; comparing the intrinsic population curves with the efficiency curves of Fig. \ref{fig:efficiency}, shows how most of such massive systems are accessible to an 8km IFO. For an aLIGO detector at full sensitivity the equivalent number of detections are around 500-10,000 yr$^{-1}$ (around 3\% of all such sources).

Figure \ref{fig:num_GW151226} shows the shows cumulative number of detections with redshift assuming that 10--10 $M_{\odot}$ BBHs dominate the source population. In this case the astronomical reach is not as impressive and so  the number of detections decrease accordingly. However, assuming our adopted rate range, we see that we could still expect around 3000--50,000 detections a year ( around 15\% of the Universal population of 10--10 $M_{\odot}$ systems) using a 8km IFO or of order 30--500 yr$^{-1}$ for aLIGO at full sensitivity.

To identify possible host galaxies, one requires detections within a small enough cosmological volume that optical surveys would have a reasonable amount of completeness; additionally, to provide smaller angular error regions closer events will have higher GW SNRs. However, the reach should be large enough that event rates are sufficient. We choose a distance of 0.4\,Gpc corresponding to the first two confident detections, GW150914 and GW151226; at this range, for a 8-km IFO, the signal SNR will be large enough to ensure that the efficiencies will be close to maximum (see Figure \ref{fig:efficiency}) and the angular resolutions will be sufficiently compact to minimise the number of galaxies.

Assuming populations of both 30--30 $M_{\odot}$ BBHs and 10--10 $M_{\odot}$ BBHs our calculations predict around 2-30 detections a year respectively within a range of 0.4\,Gpc ($z \sim 0.09$). Within 0.4\,Gpc the detection efficiencies of aLIGO are high enough that similar numbers could be expected; however, the smaller SNRs mean that the angular error regions will be much larger.

\section{Angular resolution of detector networks}
\label{sect:angular_res}

To obtain estimates of angular resolutions for BBH mergers we have followed the method provided in \citet{wen10}. This method estimates unknown parameters using the Fisher information matrix and calculates a lower bound on the parameter estimates that is method-independent. We assume that the GW waveform of BBH signals are known, and only their locations are the unknown parameters. To generate sufficient statistics, we simulate $10^3$ BBH signals at a representative distance of 0.4\,Gpc; as discussed in the previous section this distance approximately corresponds with the distance of the first two confirmed GW events GW150914 and GW151226 and is also in the range of galaxy survey completeness for follow-up of such GW sources (see section \ref{sect:host_gal}).

We simulate two synthetic populations based on the two LIGO detections assuming equal components of both 30\,$M_{\odot}$ and 10\,$M_{\odot}$. For each simulated source we randomise the polarisation, inclination, and position in the sky. The sensitivity of aLIGO (L:aLIGO-Livingston, H:aLIGO-Hanford) and AdV (V) is assumed to be at their design sensitivity; for Kagra (J) and LIGO-India (I) we assume aLIGO design sensitivity; we further model the Australian (A) and Chinese (C) detectors as 4$\times$ the sensitivity of aLIGO.

\begin{table*}
   \caption{The localisation statistics obtained for the different GW detector networks shown in the first column for 30-30 $M_{\odot}$ BBH mergers at a distance of 0.4Gpc. Based on the 90\% credible regions of our simulations we show the percentage of sources expected to be localised within different size error regions from 0.1\,deg$^{2}$ to 10\,deg$^{2}$. The individual instruments are denoted as follows: L:aLIGO-Livingston, H:aLIGO-Hanford, V:Advanced Virgo, J:Kagra, I:Indiago, A:Australian 8km, C:China 8km.}
   \centering
   \begin{tabular}{l|p{0.5cm}|c|c|c|c|c} 
   \hline
   \hline
    Network   &      & \multicolumn{5}{l}{Percentage of detections within credible region} \\
    \vspace{-2mm}              &      & & &  &  &  \\
         &           & 0.1\,deg$^{2}$ & 0.5\,deg$^{2}$ & 1\,deg$^{2}$ & 5\,deg$^{2}$ & 10\,deg$^{2}$ \\
              \hline
LHV      &  & $0.00$  &   $1.11 $  &   $7.59 $  &   $44.65 $  &   $60.50$ \\
LHVJ     &  & $0.00$  &   $16.44 $  &   $37.59 $  &   $82.45 $  &   $91.59$ \\
LHVI     &  & $0.00$  &   $8.26 $  &   $28.93 $  &   $80.74 $  &   $91.30$ \\
LHVA     &  & $1.05$  &   $45.30 $  &   $71.48 $  &   $96.07 $  &   $98.55$ \\
LHVC     &  & $0.63$  &   $32.26 $  &   $54.02 $  &   $89.71 $  &   $95.86$ \\
LHVJI    &  & $0.00$  &   $26.29 $  &   $51.49 $  &   $90.39 $  &   $96.25$ \\
LHVJIC   &  & $1.58$  &   $42.97 $  &   $66.33 $  &   $95.09 $  &   $98.53$ \\
LHVJIA   &  & $7.31$  &   $66.74 $  &   $88.01 $  &   $99.71 $  &   $99.93$ \\
LHVJIAC  &  & $25.06$  &   $81.06 $  &   $94.63 $  &   $99.90 $  &   $100.00$ \\

   \end{tabular}
   \label{tab:150914Gpc04}
\end{table*}

\begin{table*}
   \caption{As Table \ref{tab:150914Gpc04} but for for 10-10 $M_{\odot}$ BBH mergers at a distance of 0.4Gpc.}
   \centering
   \begin{tabular}{l|p{0.5cm}|c|c|c|c|c} 
   \hline
   \hline
    Network   &      & \multicolumn{5}{l}{Percentage of detections within credible region} \\
    \vspace{-2mm}              &      & & &  &  &  \\
         &           & 0.1\,deg$^{2}$ & 0.5\,deg$^{2}$ & 1\,deg$^{2}$ & 5\,deg$^{2}$ & 10\,deg$^{2}$ \\
              \hline
LHV      &  & $0.00$  &   $0.00 $  &   $0.00 $  &   $8.36 $  &   $21.63$ \\
LHVJ     &  & $0.00$  &   $0.00 $  &   $1.99 $  &   $36.75 $  &   $59.95$ \\
LHVI     &  & $0.00$  &   $0.00 $  &   $0.00 $  &   $31.36 $  &   $56.84$ \\
LHVA     &  & $0.00$  &   $1.20 $  &   $13.48 $  &   $73.06 $  &   $88.59$ \\
LHVC     &  & $0.00$  &   $0.79 $  &   $9.21 $  &   $55.24 $  &   $74.58$ \\
LHVJI    &  & $0.00$  &   $0.00 $  &   $4.57 $  &   $51.64 $  &   $73.49$ \\
LHVJIC   &  & $0.00$  &   $1.90 $  &   $14.37 $  &   $67.48 $  &   $83.24$ \\
LHVJIA   &  & $0.00$  &   $7.74 $  &   $32.27 $  &   $88.37 $  &   $96.24$ \\
LHVJIAC  &  & $0.00$  &   $27.48 $  &   $52.58 $  &   $95.20 $  &   $99.05$ \\
   \end{tabular}
   \label{tab:151226Gpc04}
\end{table*}

\begin{figure}
 \centering
  \includegraphics[scale = 0.65,origin=rl]{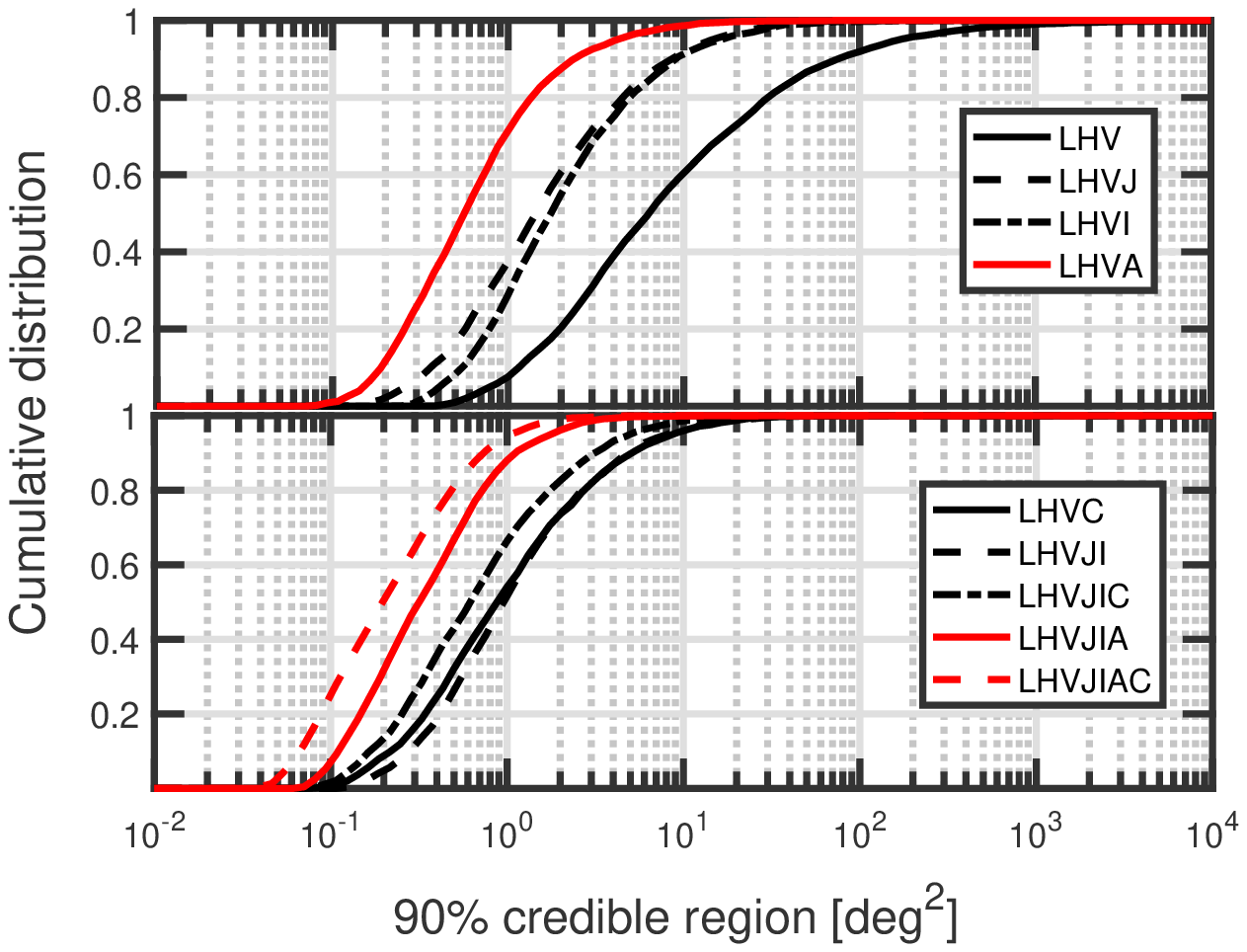}
 \caption{The localisation capabilities of different GW interferometer networks calculated for 1000 30-30 $M_{\odot}$ BBH mergers at 0.4Gpc. The source locations are randomly generated in right ascension and declination and
the polarisation and inclination angles are randomised. We denote the individual detectors in the network as follows: L:aLIGO-Livingston, H:aLIGO-Hanford, V:Advanced Virgo, J:Kagra, I:Indiago, A:Australian 8km, C:China 8km.}
 \label{fig:angres1}
\end{figure}

\begin{figure}
 \centering
   \includegraphics[scale = 0.6,origin=rl]{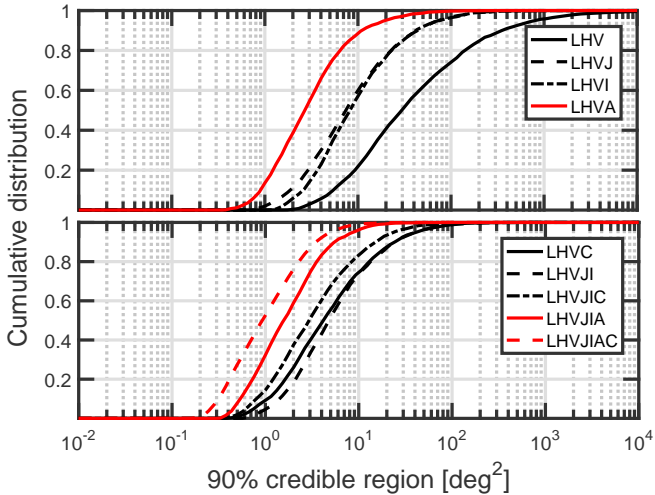}
 \caption{As for Fig.\ref{fig:angres1} but for 1000 10-10 $M_{\odot}$ BBH mergers at 0.4Gpc. }
  \label{fig:angres2}
\end{figure}

Figures \ref{fig:angres1} and \ref{fig:angres2} show the angular resolutions achievable by  a range of GW detector networks for detections at 0.4\,Gpc based on our simulations; the results are shown for synthetic populations of 30-30 $M_{\odot}$ and 10-10 $M_{\odot}$ type sources respectively. The plots show the cumulative distributions as a function of the 90\% credible regions of the angular resolution error regions: that is the percentage of sources that can be constrained within a specified localisation error region. Tables \ref{tab:150914Gpc04} and \ref{tab:151226Gpc04} summarise the main results shown in these figures for both synthetic source populations. As has been highlighted in many other studies \citep{Vitale2017PhRvD,2016arXiv161201471C, ligo_localization,fairhurst11,nissanke12,chuqi12,schutz11} the improvement in angular resolution with each new addition to the worldwide network is quite apparent.

For a population of 30-30 $M_{\odot}$ type events, starting from a LHV network we find 60\% of detections within 10\,deg$^{2}$ improving to 1\,deg$^{2}$ with the inclusion of J and I. Adding in C and A individually improve the error regions to 1\,deg$^{2}$ and 0.6\,deg$^{2}$ respectively for 50\% of the detections. For a full LHVKIAC network the error regions become of order 0.2\,deg$^{2}$ for 50\% of the sources and around 0.1\,deg$^{2}$ for nearly 30\% of the sources.

For a population of 10-10 $M_{\odot}$ type events we find 50\% of detections of a LHV network would be within 30\,deg$^{2}$ improving to 5\,deg$^{2}$ with the inclusion of J and I. Adding in C and A individually improve the error regions to 5\,deg$^{2}$ and 3\,deg$^{2}$ respectively for 50\% of the detections. For a full LHVJIAC network the error regions become of order 1\,deg$^{2}$ for 50\% of the sources and less than than 0.5\,deg$^{2}$ for 20\% of the sources.

It should be noted that the error region estimates presented here do not
include the effects of calibration uncertainty. Calibration
uncertainties are typically 10\% in strain amplitude and 10 degrees in
waveform phase for interferometric gravitational wave detectors \citep[see][for a recent example]{Abbott2017PhRvDa}. Studies have show that such
calibration uncertainties will lead to systematic errors in sky
localisation estimates \citep[e.g.][]{Vitale2012PhRvD}.

We note that impact of
calibration uncertainty is within the statistical uncertainties for SNRs
close to the detection threshold (SNR $~10$). However, for higher SNR
signals, such as the loudest 10\% of the detected signals discussed here,
calibration uncertainties will be a significant limitation. Therefore,
calibration of future generation GW detectors must be improved to allow
the required level of precision.

\section{Identification of host galaxies}
\label{sect:host_gal}
The possibility of host galaxy identification for BBH mergers depends on the ability to localise the source in three dimensions. Localisation on the sky depends on the array properties and the SNR as discussed above. The depth of localisation is determined by the luminosity distance estimate which itself depends on the SNR. For an advanced network of IFOs \citet{Ghosh_PRD_2015} have simulated the fractional errors in the luminosity distance for BBH mergers; their distributions peak at around 40\%. As the errors in luminosity distance are expected to be smaller for the loader higher SNR sources, we conservatively select a value of 30\% \footnote{for BH-NS sources, the simulations of \citet{Nissanke2013} find distance errors clustered around 15-25\%}.

\begin{figure}
 \centering
 \includegraphics[scale = 0.65,origin=rl]{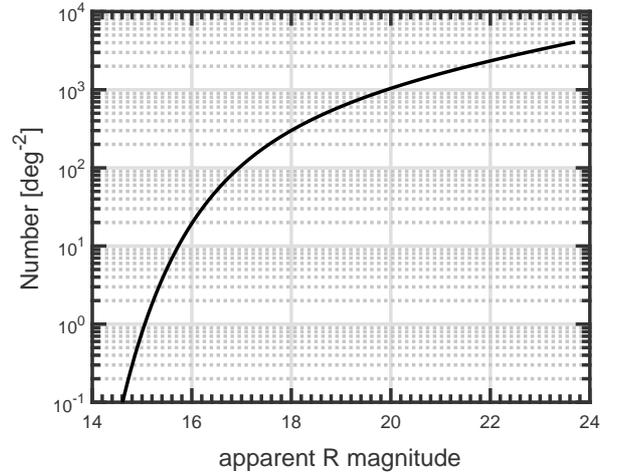}
 \caption{The integral number of galaxies at fixed distance of $400 \pm 120$ Mpc per deg$^{2}$ accessible at different survey magnitude limits. We assume an error of $\pm 30\% $ in luminosity distance for the GW source. For comparison, a Milky Way sized galaxy would have an apparent $R$ magnitude of around 17 at this distance, yielding $\sim 100$ to 200 galaxies per deg$^{2}$.}
  \label{fig:num_galaxies}
\end{figure}

We have seen above that for 30-30 $M_{\odot}$ BBH mergers at 0.4\,Gpc the angular resolution for around 30\% of the sources falls within $10^{-1}$ deg$^{2}$. Given an uncertainty in luminosity distance of $\pm 30\%$ we determine the fraction of galaxies in a volume defined by sky location and distance uncertainty, $\Delta \Theta \Delta d$ \citet[see also][]{2016ApJ...829L..15S,2016arXiv161201471C}.   
Figure \ref{fig:num_galaxies} shows predicted number counts (per deg$^{2}$ of sky) for
galaxies with distances in our BBH detection range ($d_{\mathrm{L}}=400 \pm 120$ Mpc).
Number counts were derived by integrating the $r$-band luminosity function of
\citet{Loveday_2012_MNRAS} over those luminosities corresponding to typical
apparent magnitudes detection limits ($r=14$ to 24). The Loveday et al. luminosity function
is derived from the GAMA\footnote{GAMA: Galaxy and Mass Assembly survey - for further details see \url{http://www.gama-survey.org/}} survey which probes similar redshifts
($z \sim $0.05 -- 0.25) to those in our angular resolution calculation (Sect. \ref{sect:angular_res}). No redshift evolution is implemented and similar values are
obtained using the lower redshift ($\bar{z} = 0.05$) $r$-band luminosity function of the
6dF Galaxy Survey \citep{2006MNRAS_Jones}.

Figure \ref{fig:num_galaxies} shows that around $10^2$ galaxies are observed per deg$^{2}$
brighter than $r=17$, (1 galaxy per 0.1 deg$^{2}$). This corresponds to a detecting a
single Milky Way sized galaxy at our canonical distance of 0.4 Gpc \citep[see discussion in][on the
benefits of strategies that target the brightest galaxies]{2016ApJ...820..136G}.
At deeper limits ($r=20$) the number of galaxies in the same volume increases tenfold
($10^3$ per deg$^{2}$).


Assuming a population of 30-30 $M_{\odot}$, our estimates show that a full LHVKIAC network
would localise around 30\% of the sources to within 0.1\,deg$^{2}$ (on average).
Scaling this by the estimated detection rate (Sect. \ref{sect:event_rate}) yields
$\sim 1$ to 10 events per year. In this scenario, a handful of BBH mergers could be
uniquely associated with $r \leq 17$ bright galaxies in a single year of observation.
However, for an $r \leq 20$ survey, there are $\sim 10$ galaxies per field per BBH merger,
meaning that a few years of observation would be necessary for statistical localisation.

\begin{figure}
 \centering
  \includegraphics[scale = 0.67,origin=rl]{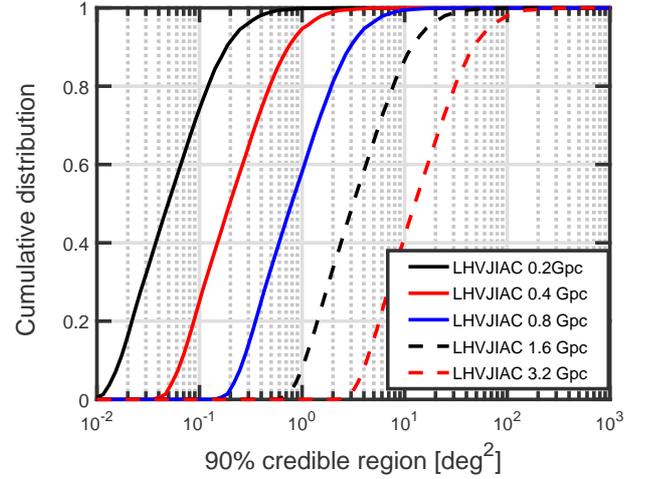}
 \caption{The localisation capabilities the largest GW interferometer network considered in this study (LHVJIAC) calculated for $10^3$ 30-30 $M_{\odot}$ BBH mergers at different distances. As previously, the source locations are randomly generated in right ascension and declination, while the polarisation and inclination angles are randomised. }
  \label{fig:SNRvsRES}
\end{figure}

As shown elsewhere in this paper (e.g. \ref{fig:horizon}), 8-km instruments will routinely
detect BBH mergers out to significantly greater distances than 0.4 Gpc. Figure \ref{fig:SNRvsRES}shows how the angular resolution (roughly proportional to the SNR) increases with astronomical distance. In the case of 30-30 $M_{\odot}$ BBH mergers at a distance of 0.8 Gpc ($z \sim 0.17$), 20\% of the events could be localised within around 0.4 deg$^{2}$; this corresponds to $\sim 10$ to 200 BBH events per year.

\begin{figure}
 \centering
 \includegraphics[scale = 0.65,origin=rl]{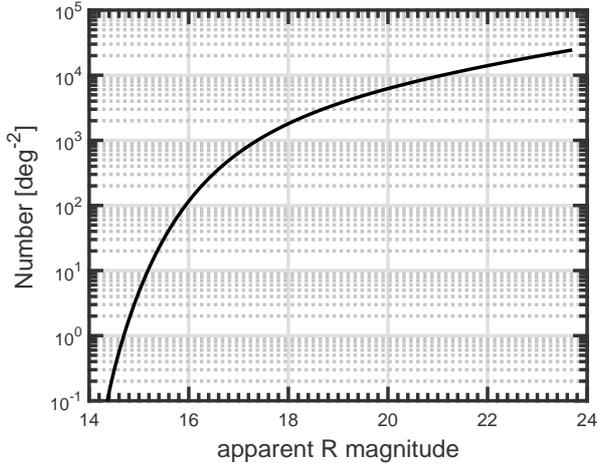}
 \caption{We repeat the calculation of Figure \ref{fig:num_galaxies} within $800 \pm 240$ Mpc per deg$^{2}$. }
  \label{fig:num_galaxies2}
\end{figure}

To estimate the number of galaxies within a 0.4 deg$^{2}$ field-of-view at
0.8\,Gpc, we repeat our galaxy number calculation. This time we determine the total number of
galaxies within the range $800 \pm 240$ Mpc ($ 0.12 \leq z < 0.21$) per deg$^{2}$ as a
function of limiting magnitude. As before, the distance range reflects the $\pm 30\%$ assumed
luminosity distance uncertainty for the GW source.

Figure \ref{fig:num_galaxies2} shows the results of this calculation. At the limit $r=17$ we
expect around 100 bright galaxies in a 0.4 deg$^{2}$ field around a 0.8\,Gpc-distant BBH merger.
As expected, the greater cosmological volumes make source confusion problematic. These issues are
compounded by the increased observational difficulty of confirming galaxy redshifts at
greater distances.

We suggest that although less frequent, the low-redshift events with high SNR are the most valuable in terms of assigning host galaxy associations. In practice, identification of host galaxies for nearby GW sources would provide a guideline in assigning likelihood rankings among the galaxies in the confused field for distant sources. For example, if the gravitational sources are predominantly found from nearby star forming galaxies, galaxies with greater blue luminosity more likely to be the hosts. In this way we can progressively widen the horizon.

\section{An 8km detector operating at low optical power}

\begin{figure}
 \centering
 \includegraphics[bbllx = 70pt,bblly =5pt, bburx = 350pt, bbury =320pt,scale = 0.6,origin=rl]{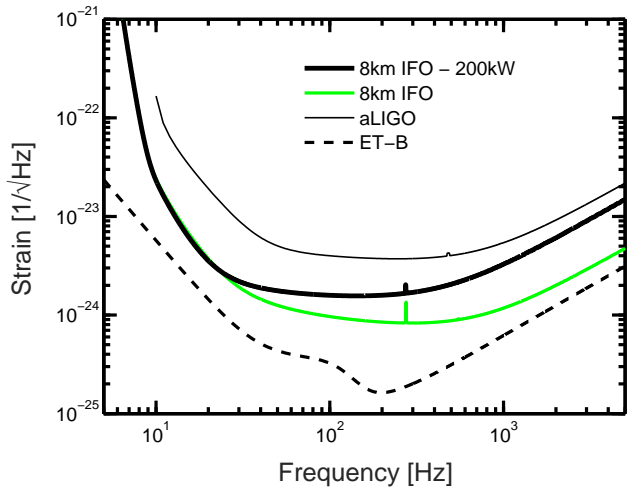}
 \caption{The sensitivity noise curve of an 8km interferometer operating with 200kW arm cavity power (8km-200kW IFO) is compared to the 1.4MW configuration (8km IFO), aLIGO at design sensitivity and a proposed configuration of the Einstein telescope, ET-B. }
  \label{fig:8kmIFO200kW}
\end{figure}

The analysis above considered the performance an 8km interferometer operating with an arm cavity power of 1.4MW. It is interesting to also consider the performance if the arm cavity power was reduced to 200kW, since high optical power is one of the greatest technical challenges. It is well known that high power introduces thermal aberrations \citep{Zhao_PhysRevLett_2006} and can cause parametric instability \citep{Zhao_PhysRevLett_2006a,Zhao_PhysRevD_2015,2015PhRvL.114p1102E}. The value of 200kW is chosen because it is comparable to the power levels already achieved in aLIGO.

Figure \ref{fig:8kmIFO200kW} shows the sensitivity noise curve of an 8km IFO operating with 200kW arm cavity power. For comparison it is compared to the 1.4MW configuration, aLIGO at design sensitivity and a proposed configuration of the Einstein telescope, ET-B \citep{2008arXiv0810.0604H}. We see that up to around 30\,Hz there no change and up to around 270\,Hz the performance is degraded by less than a factor of 2.

\begin{table}
   \caption{A comparison of the performance between 8\,km interferometric detectors operating with 1.4MW and 200kW arm cavity power. The average reach is defined as the redshift range at which the detection efficiency is 50\%.}
   \centering
   \begin{tabular}{l|l|l|l|l}
   \hline
            & \multicolumn{2}{c}{GW150914-type }  & \multicolumn{2}{c}{GW151226-type}\\
                    &     1.4MW   &     200kW   &  1.4MW       &  200kW   \\
               \hline
    Horizon redshift & 8.8  & 8.7  & 4.3 & 2.5  \\
    Average reach    &2.7  & 1.9 & 0.9 & 0.6  \\
    Detection rate (yr$^{-1}$)    & 4100  & 3100 & 8300 & 3800  \\
    \hline
   \end{tabular}
   \label{tab:8km200Mw}
\end{table}

\begin{figure}
 \centering
 \includegraphics[bbllx = 80pt,bblly =5pt, bburx = 350pt, bbury =320pt,scale = 0.6,origin=rl]{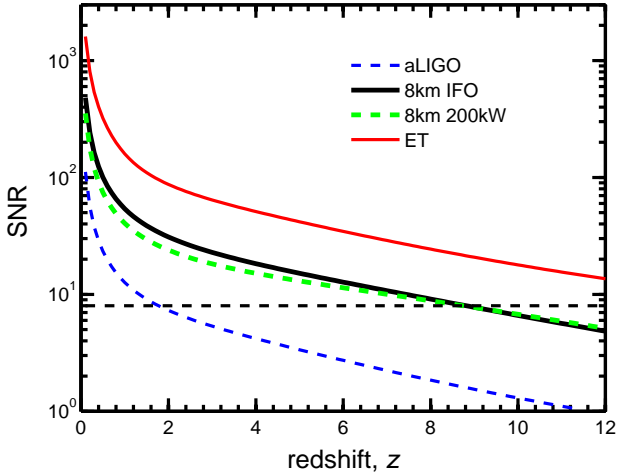}
 \caption{The optimal SNR with redshift for a GW150914 type system is shown for the nominal 8\,km detector (1.4MW optical cavity power) and for an 8\,km detector with 200kW optical cavity power. Both these instruments are compared with aLIGO and the ET-B third generation configuration. The SNR is similar at high-$z$ (due to similar low frequency performance) but differs by around a factor of nearly 2 at around z$\sim 2$.
}
   \label{fig:snr_z_GW150914}
\end{figure}

Table \ref{tab:8km200Mw} shows a comparison of the horizon distance and event rates for both 200kW and at the nominal 1.4MW arm cavity power; for reference we show results based on the first two confirmed GW detections. The loss of performance is less than might be expected because the events we are considering are dominated by low frequency noise and thermal noise which are reduced in this design; this is particularly evident for heavier GW150914 type systems. For the heavier type systems the horizon redshifts are comparable; however the average ranges (defined as the distance at which the detection efficiency is 50\%) do vary. This can be explained by Figure \ref{fig:snr_z_GW150914} which compares the optimal SNR with redshift for the two 8\,km configurations. The plot shows that due to similar low frequency performance, the SNR is comparible at high-$z$ but differs by around a factor of nearly 2 at the intermediate range of around z$\sim 2$.

The loss of sensitivity at high frequency, which scales as the square root of the power ratio would affect ability to detect black hole quasi-normal modes at larger ranges, and also high frequency sources such as binary neutron star mergers. We find for these latter sources the horizon distance decreases from 1.9\,Gpc (\emph{z}=0.36) to 1.4\,Gpc (\emph{z}=0.27) if the arm cavity power is reduced to 200kW.

\section{Discussion}

For this study we have based our modelling around the three unambiguously identified signals; GW150914, GW151226 and GW170104 using two canonical GW signals; those from 30--30 $M_{\odot}$ and 10--10 $M_{\odot}$ BBH mergers. An alternative would have been to conduct an analysis based on a model mass distribution. However, given that there still exist many uncertainties in the BBH formation channels and the intrinsic event rate from only 3 (possibly 4 including LVT151012) observations we have chosen the former option to present a case solely based on observations.  Therefore the two scenarios we consider span the possibility space based on present knowledge (heavy system/upper rate; lighter system/lower rate) to provide an estimate of the astrophysics that could be accessed by an 8\,km detector.

For a population dominated by  30--30 $M_{\odot}$ type events we find the limit of the detection horizon of order $z\sim8.8$ and a 50\% chance of detecting sources at $z\sim2.8$. One could conservatively expect between 10$^{4}$ to 10$^{5}$ detections a year.

For the less massive 10--10 $M_{\odot}$ type events we find of order 3000 - 50,000 detections in 1-year of observation. The astronomical reach to these lighter events is still impressive; a horizon distance of $z \sim 4.26$ and a 50\% probability of detection at $z \sim 0.9$. Such an astronomical reach is sufficient to capture a large proportion of events as the peak of the source distribution is around $z \sim$ 1.5--2.

Assuming a population of 30-30 $M_{\odot}$ our estimates show that a full LHVKIAC network would on average localise around 30\% of the sources to within 0.1\,deg$^{2}$; we find that this corresponds to around 1--10 BBH events per year. Our calculations suggest that if one considers the brightest galaxies of a survey, greater that r=17 in apparent magnitude, unique GW-galaxy associations may be possible. At deeper magnitudes down to r=20, one can expect 10 galaxies in a 0.1\,deg$^{2}$ field for each BBH merger leading to data of useful statistical significance. At greater distances than 0.4 Gpc, although the number of detections will be greater, a combination of lower SNR and this larger angular error regions, source confusion and survey incompleteness suggests that for an 8-km IFO, the optimal strategy to find host galaxies is to target GW sources within 0.4 Gpc. One should note however, that even at closer distances, another factor that will reduce the number of galaxy localisations is the Galactic plane, which obstructs the view of around 20\% of the extragalactic sky at visible wavelengths.

The prospect of host galaxy identification relies on the GW network's
ability to resolve sources to within 0.1 deg$^2$. The ability to achieve
the desired angular resolution is subject to the effects of calibration
uncertainties which will limit the network's sky localisation
capabilities. Therefore, the improved calibration techniques are
required for future GW detectors.

We have also considered an alternative design in which the arm cavity power is reduced from 1.4MW to 200kW, similar to the power levels already achieved in aLIGO. Our motivation here is to circumvent some of the difficulties encountered with high optical powers such as parametric instabilities. We find comparable performance levels in detecting heavier GW150914 type systems (particularly for the more distant events with greater redshifted masses) but around a factor of 2 less detections for GW151226 type systems. Therefore an 8\,km interferometric detector with reduced optical power can still provide an abundance of detections in the range 1000--100,000 events per year.

\section{Conclusion}

We have shown that a realistic design concept for one or two 8km laser interferometer gravitational wave detectors, could enable the detection of up to 3000--200,000 binary black hole mergers per year. For a small proportion (1-10) this could allow sources to directly identified with bright host galaxies. If such an association was proven, this would allow independent red-shift measurements as well as allowing detailed study of the BBH population with view to understanding their origins. At deeper magnitude limits, say $r \sim 20$, some 10s of galaxies would occupy a $\Delta \Theta \Delta d$ volume - such samples could enable host galaxy associations to be probed statistically.

An array with such high angular resolution could also be very important for mapping stochastic backgrounds, gravitational wave burst sources and binary neutron stars, as well as being a powerful tool for multi-messenger astronomy. The individual high sensitivity detectors would allow much deeper searches for continuous gravitational waves from spinning neutron stars.

Because the 8km detector is not dependent on new technologies for initial operation, it could be brought on line rapidly, thereby enabling greatly improved angular resolution and event rates in the expanding world wide network. It is important to note that in analogous Very Long Baseline Interferometry systems for high angular resolution radio astronomy, it is common for telescopes to have varying performance. Thus there would be no incompatibility issues associated with the installation of a significantly improved detector because noise contributes as the geometric mean.

\section{ACKNOWLEDGMENTS}
EJH acknowledges support from a Australian Research Council DECRA Fellowship (DE170100891). DC is supported by an Australian Research Council Future Fellowship (FT100100345). Z.-H. Zhu was supported by the National Basic Science Program (Project 973) of China (Grant No. 2014CB845800), the National Natural Science Foundation of China under Grants Nos. 11633001 and 11373014, and the Strategic Priority Research Program of the Chinese Academy of Sciences, Grant No. XDB23000000. We thank M. Colless (Australian National University) for useful discussions which aided the analysis in section 5.


\begin{thebibliography}{}

\bibitem[\protect\citeauthoryear{{Aasi}, {Abbott}, {Abbott}, {Abbott},
  {Abernathy}, {Ackley}, {Adams}, {Adams}, {Addesso} \& et al.}{{Aasi}
  et~al.}{2015}]{aLIGO_2015CQGra}
{Aasi} J.,  {Abbott} B.~P.,  {Abbott} R.,  {Abbott} T.,  {Abernathy} M.~R.,
  {Ackley} K.,  {Adams} C.,  {Adams} T.,  {Addesso} P.,    et al. 2015,
  Classical and Quantum Gravity, 32, 074001

\bibitem[\protect\citeauthoryear{Aasi et~al.,}{Aasi
  et~al.}{2013}]{ligo_localization}
Aasi J.,  et~al.,, 2013, {P}rospects for {L}ocalization of {G}ravitational
  {W}ave {T}ransients by the {A}dvanced {L}{I}{G}{O} and {A}dvanced {V}irgo
  {O}bservatories

\bibitem[\protect\citeauthoryear{{Abadie}, {Abbott}, {Abbott}, {Abernathy},
  {Accadia}, {Acernese}, {Adams}, {Adhikari}, {Ajith}, {Allen} \& et
  al.}{{Abadie} et~al.}{2010}]{Abadie2010CQG}
{Abadie} J.,  {Abbott} B.~P.,  {Abbott} R.,  {Abernathy} M.,  {Accadia} T.,
  {Acernese} F.,  {Adams} C.,  {Adhikari} R.,  {Ajith} P.,  {Allen} B.,    et
  al. 2010, Classical and Quantum Gravity, 27, 173001

\bibitem[\protect\citeauthoryear{{Abbott}, {Abbott}, {Abbott}, {Abernathy},
  {Acernese}, {Ackley}, {Adams}, {Adams}, {Addesso}, {Adhikari} \& et
  al.}{{Abbott} et~al.}{2016c}]{GW150914_AstImp_2016ApJ}
{Abbott} B.~P.,  {Abbott} R.,  {Abbott} T.~D.,  {Abernathy} M.~R.,  {Acernese}
  F.,  {Ackley} K.,  {Adams} C.,  {Adams} T.,  {Addesso} P.,  {Adhikari} R.~X.,
     et al. 2016c, \apjl, 818, L22

\bibitem[\protect\citeauthoryear{{Abbott}, {Abbott}, {Abbott}, {Abernathy},
  {Acernese}, {Ackley}, {Adams}, {Adams}, {Addesso}, {Adhikari} \& et
  al.}{{Abbott} et~al.}{2016d}]{BBH_GW150914_SGWB_2016PhRvL}
{Abbott} B.~P.,  {Abbott} R.,  {Abbott} T.~D.,  {Abernathy} M.~R.,  {Acernese}
  F.,  {Ackley} K.,  {Adams} C.,  {Adams} T.,  {Addesso} P.,  {Adhikari} R.~X.,
     et al. 2016d, Physical Review Letters, 116, 131102

\bibitem[\protect\citeauthoryear{{Abbott}, {Abbott}, {Abbott}, {Abernathy},
  {Acernese}, {Ackley}, {Adams}, {Adams}, {Addesso}, {Adhikari} \& et
  al.}{{Abbott} et~al.}{2016b}]{2016PhRvL_GW151226}
{Abbott} B.~P.,  {Abbott} R.,  {Abbott} T.~D.,  {Abernathy} M.~R.,  {Acernese}
  F.,  {Ackley} K.,  {Adams} C.,  {Adams} T.,  {Addesso} P.,  {Adhikari} R.~X.,
     et al. 2016b, Physical Review Letters, 116, 241103

\bibitem[\protect\citeauthoryear{{Abbott}, {Abbott}, {Abbott}, {Abernathy},
  {Acernese}, {Ackley}, {Adams}, {Adams}, {Addesso}, {Adhikari} \& et
  al.}{{Abbott} et~al.}{2016a}]{2016PhRvL_GW150914}
{Abbott} B.~P.,  {Abbott} R.,  {Abbott} T.~D.,  {Abernathy} M.~R.,  {Acernese}
  F.,  {Ackley} K.,  {Adams} C.,  {Adams} T.,  {Addesso} P.,  {Adhikari} R.~X.,
     et al. 2016a, Physical Review Letters, 116, 061102

\bibitem[\protect\citeauthoryear{{Abbott}, {Abbott}, {Abbott}, {Abernathy},
  {Acernese}, {Ackley}, {Adams}, {Adams} \& et al.}{{Abbott}
  et~al.}{2016}]{O1_BBHs_2016}
{Abbott} B.~P.,  {Abbott} R.,  {Abbott} T.~D.,  {Abernathy} M.~R.,  {Acernese}
  F.,  {Ackley} K.,  {Adams} C.,  {Adams} T.,    et al. 2016, preprint
  (arXiv:1606.04856)

\bibitem[\protect\citeauthoryear{{Abbott}, {Abbott}, {Abbott}, {Abernathy},
  {Acernese}, {Ackley}, {Adams}, {Adams} \& et al.}{{Abbott}
  et~al.}{2017}]{GW:170104}
{Abbott} B.~P.,  {Abbott} R.,  {Abbott} T.~D.,  {Abernathy} M.~R.,  {Acernese}
  F.,  {Ackley} K.,  {Adams} C.,  {Adams} T.,    et al. 2017, preprint
  (arXiv:xxxx.xxxx)

\bibitem[\protect\citeauthoryear{{Abbott}, {Abbott}, {Abbott}, {Abernathy},
  {Ackley}, {Adams}, {Addesso}, {Adhikari}, {Adya}, {Affeldt} \& et
  al.}{{Abbott} et~al.}{2016}]{Abbott_CE_2016}
{Abbott} B.~P.,  {Abbott} R.,  {Abbott} T.~D.,  {Abernathy} M.~R.,  {Ackley}
  K.,  {Adams} C.,  {Addesso} P.,  {Adhikari} R.~X.,  {Adya} V.~B.,  {Affeldt}
  C.,    et al. 2016, ArXiv e-prints

\bibitem[\protect\citeauthoryear{{Abbott}, {Abbott}, {Abbott}, {Abernathy},
  {Ackley}, {Adams}, {Addesso}, {Adhikari}, {Adya}, {Affeldt} \& et
  al.}{{Abbott} et~al.}{2017}]{Abbott2017PhRvDa}
{Abbott} B.~P.,  {Abbott} R.,  {Abbott} T.~D.,  {Abernathy} M.~R.,  {Ackley}
  K.,  {Adams} C.,  {Addesso} P.,  {Adhikari} R.~X.,  {Adya} V.~B.,  {Affeldt}
  C.,    et al. 2017, \prd, 95, 062003

\bibitem[\protect\citeauthoryear{Acernese et~al.,}{Acernese
  et~al.}{2015}]{TheVirgo:2014hva}
Acernese F.,  et~al., 2015, Class. Quant. Grav., 32, 024001

\bibitem[\protect\citeauthoryear{{Ajith} et~al.,}{{Ajith}
  et~al.}{2008}]{Ajith_2008}
{Ajith} P.,  et~al., 2008, \prd, 77, 104017

\bibitem[\protect\citeauthoryear{Aso, Michimura, Somiya, Ando, Miyakawa,
  Sekiguchi, Tatsumi \& Yamamoto}{Aso et~al.}{2013}]{KAGRA_PhysRevD_2013}
Aso Y.,  Michimura Y.,  Somiya K.,  Ando M.,  Miyakawa O.,  Sekiguchi T.,
  Tatsumi D.,    Yamamoto H.,  2013, Phys. Rev. D, 88, 043007

\bibitem[\protect\citeauthoryear{{Bae}, {Kim} \& {Lee}}{{Bae}
  et~al.}{2014}]{Bae2014MNRAS}
{Bae} Y.-B.,  {Kim} C.,    {Lee} H.~M.,  2014, \mnras, 440, 2714

\bibitem[\protect\citeauthoryear{{Belczynski}, {Buonanno}, {Cantiello},
  {Fryer}, {Holz}, {Mandel}, {Miller} \& {Walczak}}{{Belczynski}
  et~al.}{2014}]{Belczynski2014ApJ}
{Belczynski} K.,  {Buonanno} A.,  {Cantiello} M.,  {Fryer} C.~L.,  {Holz}
  D.~E.,  {Mandel} I.,  {Miller} M.~C.,    {Walczak} M.,  2014, \apj, 789, 120

\bibitem[\protect\citeauthoryear{{Belczynski}, {Holz}, {Bulik} \& {O
  Shaughnessy}}{{Belczynski} et~al.}{2016}]{Belczynski_2016_Natur}
{Belczynski} K.,  {Holz} D.~E.,  {Bulik} T.,    {O Shaughnessy} R.,  2016,
  \nat, 534, 512

\bibitem[\protect\citeauthoryear{{Blair}, {Ju}, {Zhao}, {Wen}, {Miao}, {Cai},
  {Gao}, {Lin}, {Liu}, {Wu}, {Zhu}, {Hammond}, {Paik}, {Fafone}, {Rocchi},
  {Blair}, {Ma}, {Qin} \& {Page}}{{Blair} et~al.}{2015}]{Blair2015SCPMA}
{Blair} D.,  {Ju} L.,  {Zhao} C.,  {Wen} L.,  {Miao} H.,  {Cai} R.,  {Gao} J.,
  {Lin} X.,  {Liu} D.,  {Wu} L.-A.,  {Zhu} Z.,  {Hammond} G.,  {Paik} H.~J.,
  {Fafone} V.,  {Rocchi} A.,  {Blair} C.,  {Ma} Y.,  {Qin} J.,    {Page} M.,
  2015, Science China Physics, Mechanics, and Astronomy, 58, 5747

\bibitem[\protect\citeauthoryear{{Chen} \& {Holz}}{{Chen} \&
  {Holz}}{2016}]{2016arXiv161201471C}
{Chen} H.-Y.,  {Holz} D.~E.,  2016, ArXiv e-prints

\bibitem[\protect\citeauthoryear{{Chu}, {Howell}, {Rowlinson}, {Gao}, {Zhang},
  {Tingay}, {Bo{\"e}r} \& {Wen}}{{Chu} et~al.}{2016}]{Chu2016MNRAS}
{Chu} Q.,  {Howell} E.~J.,  {Rowlinson} A.,  {Gao} H.,  {Zhang} B.,  {Tingay}
  S.~J.,  {Bo{\"e}r} M.,    {Wen} L.,  2016, \mnras, 459, 121

\bibitem[\protect\citeauthoryear{Chu, Wen \& Blair}{Chu et~al.}{2012}]{chuqi12}
Chu Q.,  Wen L.,    Blair D.,  2012, Journal of Physics: Conference Series,
  363, 012023

\bibitem[\protect\citeauthoryear{Coughlin, Mukund, Harms, Driggers, Adhikari \&
  Mitra}{Coughlin et~al.}{2016}]{0264-9381-33-24-244001}
Coughlin M.,  Mukund N.,  Harms J.,  Driggers J.,  Adhikari R.,    Mitra S.,
  2016, Classical and Quantum Gravity, 33, 244001

\bibitem[\protect\citeauthoryear{{Dominik}, {Berti}, {OShaughnessy}, {Mandel},
  {Belczynski}, {Fryer}, {Holz}, {Bulik} \& {Pannarale}}{{Dominik}
  et~al.}{2015}]{Dominik2015ApJ}
{Dominik} M.,  {Berti} E.,  {OShaughnessy} R.,  {Mandel} I.,  {Belczynski} K.,
  {Fryer} C.,  {Holz} D.~E.,  {Bulik} T.,    {Pannarale} F.,  2015, \apj, 806,
  263

\bibitem[\protect\citeauthoryear{{Dwyer}, {Sigg}, {Ballmer}, {Barsotti},
  {Mavalvala} \& {Evans}}{{Dwyer} et~al.}{2015}]{Dwyer2015PhRvD}
{Dwyer} S.,  {Sigg} D.,  {Ballmer} S.~W.,  {Barsotti} L.,  {Mavalvala} N.,
  {Evans} M.,  2015, \prd, 91, 082001

\bibitem[\protect\citeauthoryear{{Evans} et~al.,}{{Evans}
  et~al.}{2015}]{2015PhRvL.114p1102E}
{Evans} M.,  et~al., 2015, Physical Review Letters, 114, 161102

\bibitem[\protect\citeauthoryear{{Fairhurst}}{{Fairhurst}}{2011}]{fairhurst11}
{Fairhurst} S.,  2011, Classical and Quantum Gravity, 28, 105021

\bibitem[\protect\citeauthoryear{{Fan}, {Messenger} \& {Heng}}{{Fan}
  et~al.}{2014}]{Fan2014ApJ}
{Fan} X.,  {Messenger} C.,    {Heng} I.~S.,  2014, \apj, 795, 43

\bibitem[\protect\citeauthoryear{{Finn}}{{Finn}}{1996}]{Finn1996PhRvD}
{Finn} L.~S.,  1996, \prd, 53, 2878

\bibitem[\protect\citeauthoryear{{Finn} \& {Chernoff}}{{Finn} \&
  {Chernoff}}{1993}]{FinnChernoff1993PhRD}
{Finn} L.~S.,  {Chernoff} D.~F.,  1993, \prd, 47, 2198

\bibitem[\protect\citeauthoryear{{Gehrels}, {Cannizzo}, {Kanner}, {Kasliwal},
  {Nissanke} \& {Singer}}{{Gehrels} et~al.}{2016}]{2016ApJ...820..136G}
{Gehrels} N.,  {Cannizzo} J.~K.,  {Kanner} J.,  {Kasliwal} M.~M.,  {Nissanke}
  S.,    {Singer} L.~P.,  2016, \apj, 820, 136

\bibitem[\protect\citeauthoryear{{Ghosh}, {Del Pozzo} \& {Ajith}}{{Ghosh}
  et~al.}{2015}]{Ghosh_PRD_2015}
{Ghosh} A.,  {Del Pozzo} W.,    {Ajith} P.,  2015, ArXiv e-prints

\bibitem[\protect\citeauthoryear{Goda, Miyakawa, Mikhailov, Saraf, Adhikari,
  McKenzie, Ward, Vass, Weinstein \& Mavalvala}{Goda et~al.}{2008}]{Goda2008}
Goda K.,  Miyakawa O.,  Mikhailov E.~E.,  Saraf S.,  Adhikari R.,  McKenzie K.,
   Ward R.,  Vass S.,  Weinstein A.~J.,    Mavalvala N.,  2008, Nat Phys, 4,
  472

\bibitem[\protect\citeauthoryear{Harms \& Paik}{Harms \&
  Paik}{2015}]{PhysRevD.92.022001}
Harms J.,  Paik H.~J.,  2015, Phys. Rev. D, 92, 022001

\bibitem[\protect\citeauthoryear{Harms \& Venkateswara}{Harms \&
  Venkateswara}{2016}]{0264-9381-33-23-234001}
Harms J.,  Venkateswara K.,  2016, Classical and Quantum Gravity, 33, 234001

\bibitem[\protect\citeauthoryear{{Hild}, {Chelkowski} \& {Freise}}{{Hild}
  et~al.}{2008}]{2008arXiv0810.0604H}
{Hild} S.,  {Chelkowski} S.,    {Freise} A.,  2008, ArXiv e-prints

\bibitem[\protect\citeauthoryear{{Hong} \& {Lee}}{{Hong} \&
  {Lee}}{2015}]{Hong_2015MNRAS}
{Hong} J.,  {Lee} H.~M.,  2015, \mnras, 448, 754

\bibitem[\protect\citeauthoryear{J. et~al.,}{J.  et~al.}{2013}]{AasiJ.2013}
J. A.,  et~al., 2013, Nat Photon, 7, 613

\bibitem[\protect\citeauthoryear{{Jones}, {Peterson}, {Colless} \&
  {Saunders}}{{Jones} et~al.}{2006}]{2006MNRAS_Jones}
{Jones} D.~H.,  {Peterson} B.~A.,  {Colless} M.,    {Saunders} W.,  2006,
  \mnras, 369, 25

\bibitem[\protect\citeauthoryear{Kimble, Levin, Matsko, Thorne \&
  Vyatchanin}{Kimble et~al.}{2001}]{PhysRevD.65.022002}
Kimble H.~J.,  Levin Y.,  Matsko A.~B.,  Thorne K.~S.,    Vyatchanin S.~P.,
  2001, Phys. Rev. D, 65, 022002

\bibitem[\protect\citeauthoryear{{Klimenko}, {Vedovato}, {Drago}, {Mazzolo},
  {Mitselmakher}, {Pankow}, {Prodi}, {Re}, {Salemi} \& {Yakushin}}{{Klimenko}
  et~al.}{2011}]{Klimenko2011PhRvD}
{Klimenko} S.,  {Vedovato} G.,  {Drago} M.,  {Mazzolo} G.,  {Mitselmakher} G.,
  {Pankow} C.,  {Prodi} G.,  {Re} V.,  {Salemi} F.,    {Yakushin} I.,  2011,
  \prd, 83, 102001

\bibitem[\protect\citeauthoryear{Langer \& Norman}{Langer \&
  Norman}{2006}]{langerNorman_metalisity_06}
Langer N.,  Norman C.~A.,  2006, ApJL, 638, L63

\bibitem[\protect\citeauthoryear{{Lipunov}, {Postnov} \& {Prokhorov}}{{Lipunov}
  et~al.}{1997}]{Lipunov1997}
{Lipunov} V.~M.,  {Postnov} K.~A.,    {Prokhorov} M.~E.,  1997,
  astro-ph/9610016, 2, 43

\bibitem[\protect\citeauthoryear{{Loveday} et~al.,}{{Loveday}
  et~al.}{2012}]{Loveday_2012_MNRAS}
{Loveday} J.,  et~al., 2012, \mnras, 420, 1239

\bibitem[\protect\citeauthoryear{{Madau} \& {Dickinson}}{{Madau} \&
  {Dickinson}}{2014}]{MadauDickinsonSFR2014ARA&A}
{Madau} P.,  {Dickinson} M.,  2014, \araa, 52, 415

\bibitem[\protect\citeauthoryear{{Messenger}, {Takami}, {Gossan}, {Rezzolla} \&
  {Sathyaprakash}}{{Messenger} et~al.}{2014}]{Messenger_2014}
{Messenger} C.,  {Takami} K.,  {Gossan} S.,  {Rezzolla} L.,    {Sathyaprakash}
  B.~S.,  2014, Physical Review X, 4, 041004

\bibitem[\protect\citeauthoryear{Mitrofanov, Chao, Pan, Kuo, Cole, Degallaix \&
  Willke}{Mitrofanov et~al.}{2015}]{Mitrofanov_2015}
Mitrofanov V.~P.,  Chao S.,  Pan H.-W.,  Kuo L.-C.,  Cole G.,  Degallaix J.,
  Willke B.,  2015, Science China Physics, Mechanics {\&} Astronomy, 58, 1

\bibitem[\protect\citeauthoryear{{Nissanke}, {Holz}, {Dalal}, {Hughes},
  {Sievers} \& {Hirata}}{{Nissanke} et~al.}{2013}]{Nissanke2013}
{Nissanke} S.,  {Holz} D.~E.,  {Dalal} N.,  {Hughes} S.~A.,  {Sievers} J.~L.,
   {Hirata} C.~M.,  2013, ArXiv e-prints

\bibitem[\protect\citeauthoryear{Nissanke, Kasliwal \& Georgieva}{Nissanke
  et~al.}{2012}]{nissanke12}
Nissanke S.,  Kasliwal M.,    Georgieva A., , 2012, Astrophysical Journal, 767,
  124 (2013)

\bibitem[\protect\citeauthoryear{Oelker, Isogai, Miller, Tse, Barsotti,
  Mavalvala \& Evans}{Oelker et~al.}{2016}]{PhysRevLett.116.041102}
Oelker E.,  Isogai T.,  Miller J.,  Tse M.,  Barsotti L.,  Mavalvala N.,
  Evans M.,  2016, Phys. Rev. Lett., 116, 041102

\bibitem[\protect\citeauthoryear{Peebles}{Peebles}{1993}]{Peebles}
Peebles P. J.~E.,  1993, Principles of Physical Cosmology, first edn.
Princeton Univ. Press, Princeton NJ

\bibitem[\protect\citeauthoryear{Pinard, Michel, Sassolas, Balzarini,
  Degallaix, Dolique, Flaminio, Forest, Granata, Lagrange, Straniero, Teillon
  \& Cagnoli}{Pinard et~al.}{2016a}]{Pinard:16}
Pinard L.,  Michel C.,  Sassolas B.,  Balzarini L.,  Degallaix J.,  Dolique J.,
   Flaminio R.,  Forest D.,  Granata M.,  Lagrange B.,  Straniero N.,  Teillon
  J.,    Cagnoli G.,  2016a, in Optical Interference Coatings 2016 The mirrors
  used in the ligo interferometers for the first-time detection of
  gravitational waves.
Optical Society of America, p.~MB.3

\bibitem[\protect\citeauthoryear{Pinard, Michel, Sassolas, Balzarini,
  Degallaix, Dolique, Flaminio, Forest, Granata, Lagrange, Straniero, Teillon
  \& Cagnoli}{Pinard et~al.}{2016b}]{Pinard_2016_IBS}
Pinard L.,  Michel C.,  Sassolas B.,  Balzarini L.,  Degallaix J.,  Dolique J.,
   Flaminio R.,  Forest D.,  Granata M.,  Lagrange B.,  Straniero N.,  Teillon
  J.,    Cagnoli G.,  2016b, Optical Interference Coatings 2016, p.~MB.3

\bibitem[\protect\citeauthoryear{{Planck Collaboration}, {Ade}, {Aghanim},
  {Arnaud}, {Ashdown}, {Aumont}, {Baccigalupi}, {Banday}, {Barreiro},
  {Bartlett} \& et al.}{{Planck Collaboration} et~al.}{2015}]{Planck2015}
{Planck Collaboration} {Ade} P.~A.~R.,  {Aghanim} N.,  {Arnaud} M.,  {Ashdown}
  M.,  {Aumont} J.,  {Baccigalupi} C.,  {Banday} A.~J.,  {Barreiro} R.~B.,
  {Bartlett} J.~G.,    et al. 2015, ArXiv e-prints

\bibitem[\protect\citeauthoryear{Punturo et~al.,}{Punturo
  et~al.}{2010}]{Punturo_ET_2010}
Punturo M.,  et~al., 2010, Classical and Quantum Gravity, 27, 084007

\bibitem[\protect\citeauthoryear{{Raffai}, {Gond{\'a}n}, {Heng},
  {Kelecs{\'e}nyi}, {Logue}, {M{\'a}rka} \& {M{\'a}rka}}{{Raffai}
  et~al.}{2013}]{Raffai2013CQGra}
{Raffai} P.,  {Gond{\'a}n} L.,  {Heng} I.~S.,  {Kelecs{\'e}nyi} N.,  {Logue}
  J.,  {M{\'a}rka} Z.,    {M{\'a}rka} S.,  2013, Classical and Quantum Gravity,
  30, 155004

\bibitem[\protect\citeauthoryear{{Regimbau} \& {Hughes}}{{Regimbau} \&
  {Hughes}}{2009}]{Regimbau2009PhRvD}
{Regimbau} T.,  {Hughes} S.~A.,  2009, \prd, 79, 062002

\bibitem[\protect\citeauthoryear{{Rodriguez}, {Chatterjee} \&
  {Rasio}}{{Rodriguez} et~al.}{2016}]{Rodriguez2016}
{Rodriguez} C.~L.,  {Chatterjee} S.,    {Rasio} F.~A.,  2016, \prd, 93, 084029

\bibitem[\protect\citeauthoryear{{Rodriguez}, {Haster}, {Chatterjee},
  {Kalogera} \& {Rasio}}{{Rodriguez} et~al.}{2016}]{Rodriguez2016ApJ}
{Rodriguez} C.~L.,  {Haster} C.-J.,  {Chatterjee} S.,  {Kalogera} V.,
  {Rasio} F.~A.,  2016, \apjl, 824, L8

\bibitem[\protect\citeauthoryear{{Rodriguez}, {Morscher}, {Pattabiraman},
  {Chatterjee}, {Haster} \& {Rasio}}{{Rodriguez} et~al.}{2015}]{Rodriguez2015}
{Rodriguez} C.~L.,  {Morscher} M.,  {Pattabiraman} B.,  {Chatterjee} S.,
  {Haster} C.-J.,    {Rasio} F.~A.,  2015, Physical Review Letters, 115, 051101

\bibitem[\protect\citeauthoryear{Schutz}{Schutz}{2011}]{schutz11}
Schutz B.~F.,  2011, Classical and Quantum Gravity, 28, 125023

\bibitem[\protect\citeauthoryear{{Singer}, {Chen}, {Holz}, {Farr}, {Price},
  {Raymond}, {Cenko}, {Gehrels}, {Cannizzo}, {Kasliwal}, {Nissanke},
  {Coughlin}, {Farr}, {Urban}, {Vitale}, {Veitch}, {Graff}, {Berry},
  {Mohapatra} \& {Mandel}}{{Singer} et~al.}{2016}]{2016ApJ...829L..15S}
{Singer} L.~P.,  {Chen} H.-Y.,  {Holz} D.~E.,  {Farr} W.~M.,  {Price} L.~R.,
  {Raymond} V.,  {Cenko} S.~B.,  {Gehrels} N.,  {Cannizzo} J.,  {Kasliwal}
  M.~M.,  {Nissanke} S.,  {Coughlin} M.,  {Farr} B.,  {Urban} A.~L.,  {Vitale}
  S.,  {Veitch} J.,  {Graff} P.,  {Berry} C.~P.~L.,  {Mohapatra} S.,
  {Mandel} I.,  2016, \apjl, 829, L15

\bibitem[\protect\citeauthoryear{{Stevenson}, {Ohme} \&
  {Fairhurst}}{{Stevenson} et~al.}{2015}]{Stevenson_2015ApJ}
{Stevenson} S.,  {Ohme} F.,    {Fairhurst} S.,  2015, \apj, 810, 58

\bibitem[\protect\citeauthoryear{{Vitale}}{{Vitale}}{2016}]{Vitale2016PhRvD}
{Vitale} S.,  2016, \prd, 94, 121501

\bibitem[\protect\citeauthoryear{{Vitale}, {Del Pozzo}, {Li}, {Van Den Broeck},
  {Mandel}, {Aylott} \& {Veitch}}{{Vitale} et~al.}{2012}]{Vitale2012PhRvD}
{Vitale} S.,  {Del Pozzo} W.,  {Li} T.~G.~F.,  {Van Den Broeck} C.,  {Mandel}
  I.,  {Aylott} B.,    {Veitch} J.,  2012, \prd, 85, 064034

\bibitem[\protect\citeauthoryear{{Vitale} \& {Evans}}{{Vitale} \&
  {Evans}}{2017}]{Vitale2017PhRvD}
{Vitale} S.,  {Evans} M.,  2017, \prd, 95, 064052

\bibitem[\protect\citeauthoryear{{Wen} \& {Chen}}{{Wen} \&
  {Chen}}{2010}]{WenChen2010PhRvD}
{Wen} L.,  {Chen} Y.,  2010, \prd, 81, 082001

\bibitem[\protect\citeauthoryear{Wen \& Chen}{Wen \& Chen}{2010}]{wen10}
Wen L.,  Chen Y.,  2010, Phys. Rev. D, 81, 082001

\bibitem[\protect\citeauthoryear{Zhao, Degallaix, Ju, Fan, Blair, Slagmolen,
  Gray, Lowry, McClelland, Hosken, Mudge, Brooks, Munch, Veitch, Barton \&
  Billingsley}{Zhao et~al.}{2006}]{Zhao_PhysRevLett_2006}
Zhao C.,  Degallaix J.,  Ju L.,  Fan Y.,  Blair D.~G.,  Slagmolen B. J.~J.,
  Gray M.~B.,  Lowry C. M.~M.,  McClelland D.~E.,  Hosken D.~J.,  Mudge D.,
  Brooks A.,  Munch J.,  Veitch P.~J.,  Barton M.~A.,    Billingsley G.,  2006,
  Phys. Rev. Lett., 96, 231101

\bibitem[\protect\citeauthoryear{Zhao, Ju, Degallaix, Gras \& Blair}{Zhao
  et~al.}{2005}]{Zhao_PhysRevLett_2006a}
Zhao C.,  Ju L.,  Degallaix J.,  Gras S.,    Blair D.~G.,  2005, Phys. Rev.
  Lett., 94, 121102

\bibitem[\protect\citeauthoryear{Zhao, Ju, Fang, Blair, Qin, Blair, Degallaix
  \& Yamamoto}{Zhao et~al.}{2015}]{Zhao_PhysRevD_2015}
Zhao C.,  Ju L.,  Fang Q.,  Blair C.,  Qin J.,  Blair D.,  Degallaix J.,
  Yamamoto H.,  2015, Phys. Rev. D, 91, 092001

\bibitem[\protect\citeauthoryear{{Zhu}, {Howell}, {Regimbau}, {Blair} \&
  {Zhu}}{{Zhu} et~al.}{2011}]{Zhu_BBH_SGWB_2011ApJ}
{Zhu} X.-J.,  {Howell} E.,  {Regimbau} T.,  {Blair} D.,    {Zhu} Z.-H.,  2011,
  \apj, 739, 86

\end{thebibliography}

\end{document}